\documentclass[5pt,two column]{article}
\usepackage{graphicx}   
\usepackage{multirow}
\usepackage{subfigure}
\usepackage{geometry}
\usepackage{stfloats}
\usepackage{booktabs}
\usepackage{amsfonts}
\usepackage{amsmath}

\usepackage{titlesec}
\titleformat{\section}
{\raggedright\normalsize\bfseries}
{\thesection .\quad}
{0pt}
{}

\titleformat{\subsection}
{\raggedright\small\bfseries}
{\thesection .\quad}
{0pt}
{}

\usepackage{caption}
\usepackage{mathrsfs}
\begin{document}\small

\captionsetup{font={small}}

\title{Flow Rate Control in Smart District Heating Systems Using Deep Reinforcement Learning}

\author{Tinghao Zhang\quad Jing Luo\quad Ping Chen\quad Jie Liu\\
Harbin Institute of Technology\\
Harbin, Heilongjiang, China\\
}
\date{}
\maketitle

\begin{abstract}

At high latitudes, many cities adopt a centralized heating system to improve the energy generation efficiency and to reduce pollution. In multi-tier systems, so-called district heating, there are a few efficient approaches for the flow rate control during the heating process. In this paper, we describe the theoretical methods to solve this problem by deep reinforcement learning and propose a cloud-based heating control system for implementation. A real-world case study shows the effectiveness and practicability of the proposed system controlled by humans, and the simulated experiments for deep reinforcement learning show about 1985.01 gigajoules of heat quantity and 42276.45 tons of water are saved per hour compared with manual control. 

\end{abstract}

\section{Introduction}
In many high-latitude areas, heating costs in winters are a significant part of the total energy consumption. For example, in Nordic countries, heating accounts for more than 60\% of all the energy use in buildings\cite{Fazeli16}. In northern China, the total heating space exceeded 9 billion square meters in 2019, which caused high heating energy consumption. Many countries are looking for better ways for heating system design.

District heating system (DHS) is widely used in many high-latitude areas, as it has higher fuel use efficiency and generates less environmental pollution. Researchers have been working on improving the DHS efficiency in various ways. Some studies have focused on optimization approaches for DHS design, which offer reasonable plans with the purpose of saving more energy and reducing costs\cite{Bordin16,Best18}. Moreover, economic analyses for DHS from different respects have been presented to ensure that DHS can work appropriately under any situation as well as minimize the costs\cite{Takacs16,Nussbaumer16}. As the source of water temperature in most DHSs is domestic hot water (DHW), some researchers have proposed several novel methods to prepare DHW at a relatively high efficiency\cite{stergaard16,Yang16}. For the control approaches, some researchers have controlled the district power consumption according to the users comfort preferences, minimizing the power consumption and costs with the help of mathematical models\cite{Choi10,Bhattacharya17}. The proportional integral differential (PID) algorithm and its improved versions have also been used in combination with other computing technologies\cite{Yang06,Lin11}. Some control models have been built using a fuzzy inference system (FIS) and an artificial neural network (ANN) to achieve the control precision and low energy costs\cite{Zhang10,Strunsik15}. With model predictive control with many system dynamics and constraints, the aggregated demand and supply water temperature was predicted, and the related algorithms controlled the district heating power plants on the basis of the predicted result\cite{Verrilli16,Verrilli17}. However, there are still some limitations for the current control means. For one thing, most of the literature has focused on the regulation of power plants, also called the heat source of the DHS, and neglected to control the flow rates. Many advanced and feasible methods have also been proposed to control the heat generation, and there is still a great deal of water wastage in the DHS, leading to considerable unnecessary costs. For another, mathematical models for optimal control or statistical models to forecast load demands cannot be universally applied because of the DHS's complexity. Different DHSs have different pipelines, scale, and connections with a large number of units. Furthermore, the heat loss from the pipes to the outside and the heat transfer efficiency differ between DHSs. Traditional models have failed to capture all the details to solve the optimization problems. More efficient control methods need to be adopted for the DHS while taking these two problems into consideration. 

Furthermore, with the fast development of the Internet of Things (IoT), Smart City\cite{Su11,Zeiger14}, as a novel concept, has attracted considerable attention. Recently, many researchers have begun to introduce this concept into DHS regulation. Unlike the traditional DHS, IoT provides operators with the opportunities to monitor the DHS in real time and offer remote monitor control, which is a more convenient and effective method for regulation. Different systems have been designed for IoT services and adopted into the DHS with the consideration of both the costs and the feasibility\cite{Sun13,Lynggaard16,Zhao16}. 

In this study, we resorted to deep reinforcement learning (DRL) for the DHS's flow rates control. Reinforcement learning (RL) has been proven to be efficient for optimal control problems\cite{Sutton98}. DRL combines the merits between deep learning (DL) and RL\cite{Mnih15} and has been successfully applied to many research areas such as robotics\cite{Vecerik17,Bruin16}, autonomous driving\cite{Sallab17,Liaw17}, and decision making\cite{Li10,Kazak19}. We built a deep neural net model for a typical DHS, and then, DRL was used to tune the flow rates. The entire process successfully achieves the purpose of optimization control without requiring a complicated mathematical representation. The method used to model the DHS in this study can be applied to other DHSs because of its universality and simplicity. Moreover, we developed a heating control system for implementation. In particular, the sensors recorded the water temperature and flow rates, and the data were sent to a cloud service through the narrow-band Internet of Things (NB-IoT). On the cloud side, operators can offer remote centralized control. The DRL algorithm will be in the cloud to achieve automatic control in the future. We also used the proposed heating control system to deal with hydraulic imbalance problems in the apartment by a PID algorithm. Our evaluation showed that the DRL algorithm managed to save approximately 1985.01 gigajoules of heat quantity and 42276.45 tons of water per hour, compared with the result controlled by humans.

\section{Heating Control Problem Formulation}
\subsection{District Heating System}
\begin{figure}[htb]
  \centering
  \includegraphics[width=8cm]{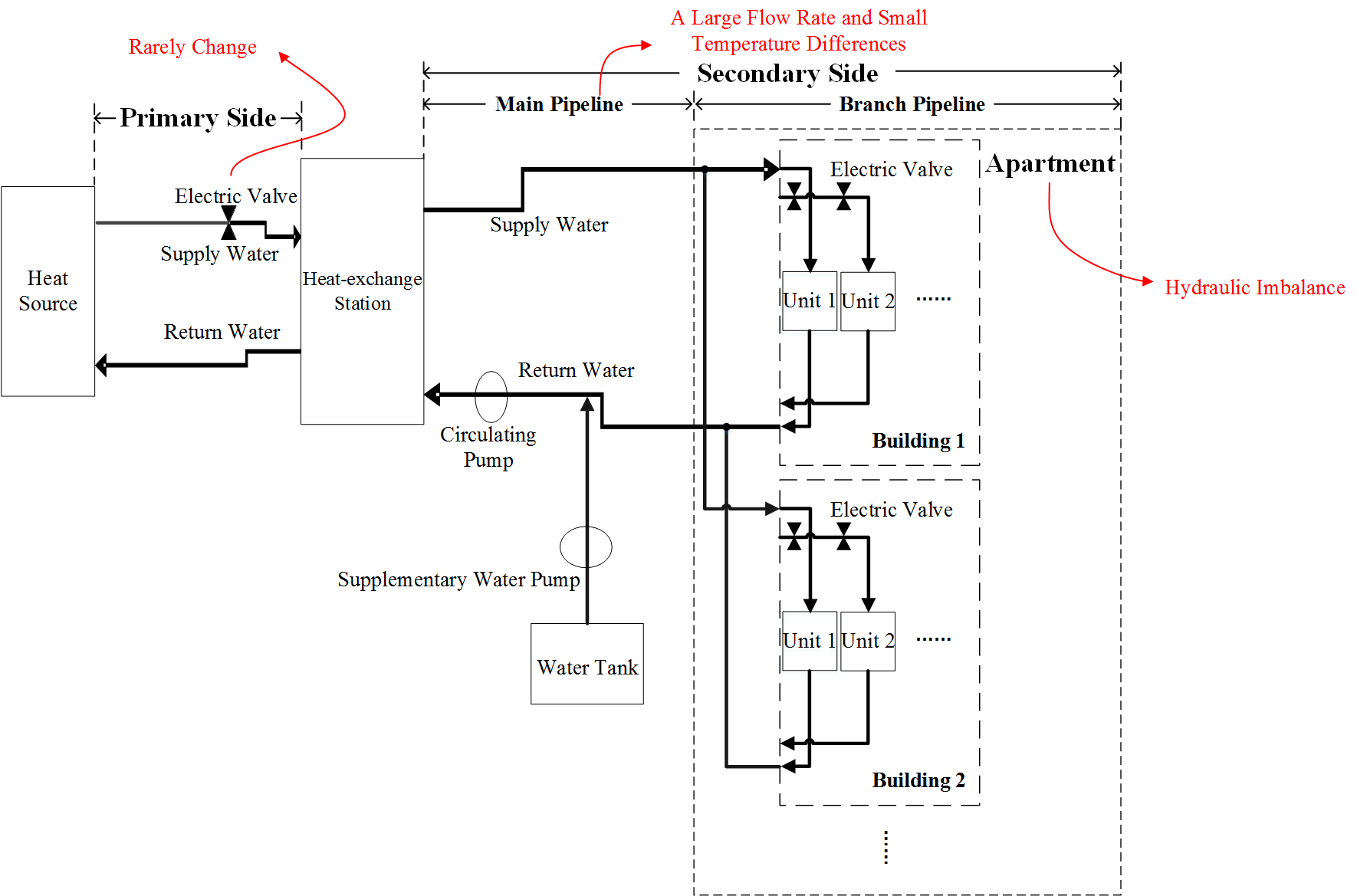}
\caption{Typical District Heating System}
\label{problem}
\end{figure}

Figure~\ref{problem} shows a typical setup for the district heating system. There are two hot water loops connected by heat-exchange stations\cite{Bhattacharya17}. The primary side of a station has the heat source and a hot water pipe that does not connect to the end users. The secondary side includes a branch pipeline in the apartment and a main pipeline. Water in the main pipeline gets the heat from the primary side through the exchange station and delivers it to individual consumers in the apartment. Pipes from each side can be divided into two classes: one is to transfer the supply water, and the other is to transfer the return water. Note that the high temperature water from the heat source has sufficient pressure, so the pipeline on the primary side does not need the addition of a pump. In comparison, a circulating pump is installed on the secondary side according to the pressure. This pump controls the total flow rate of the secondary side. In addition, because the supply and return water pipes on the secondary side sometimes leak for various reasons, which makes the pressure drop, a supplementary water pump is installed in the return pipeline of the secondary side. The supplementary water pump can supply water for the return pipes.

\subsection{Problem Statement}
There are three main challenges of DHSs, as shown in Figure~\ref{problem}. First, most of the references aimed to regulate the heat source to generate different amounts of heat energy but have seldom discussed how to tune the flow rates of both sides. Second, hydraulic imbalance\cite{Ashfaq18} is a common problem, which means that the real flow rates in the branch pipes are different from their design flow rate. The reason of hydraulic imbalance is that the distances between buildings and the heat exchange station are different. In terms of buildings that are close to the station, they sometimes obtain an excessive flow rate because of the considerable initial pressure. However, after the hot water in the branch pipeline runs through these buildings, the remaining pressure fails to send sufficient hot water to the other buildings that are far away from the station. In this case, a thermo-imbalance usually happens in household terminals as a result, which leads to uneven cold and warm room temperatures. People with a high indoor temperature tend to open windows to dissipate the heat, which is an energy waste, while those with a low indoor temperature complain about being uncomfortable. Third, with the intent of solving hydraulic imbalance problems, some community-scale heating systems operate with a large flow rate and small temperature differences\cite{Jian15,Zhang16}, with the hope that the water temperature in the branch pipes are more even. However, a fast water flow uses excessive energy to maintain the water circulation and loses a considerable amount of water. The balancing of water temperatures and the circulation speed is not trivial for human operators.

On the basis of the above discussion, we first applied the PID algorithm to obtain even flow rates on the apartment side in order to solve hydraulic imbalance problems. Therefore, the flow rate in the branch pipeline was uniform distributed even with a related small flow rate in the main pipeline. Then, for overall control, we analyzed the heat transfer process. For a typical district heating system, the hot water from a heat source (usually a boiler house) flows through the heat-exchange station as the supply water, transfers its heat to the cold water in the station, and then returns to the heat source. The supply water on the secondary side receives heat and transfers the heat to the apartments\cite{Cao19}. During the process, the temperature difference (TD) between the supply water and the return water and the transferred heat quantity $Q$ were calculated using equation~\ref{1} and equation~\ref{2}, respectively.

\begin{equation}\label{1}
    TD = T_{supply} - T_{return}
\end{equation}
\begin{equation}\label{2}
    Q = c \times F \times TD 
\end{equation}
where $c$ denotes the specific heat capacity of water and $F$ denotes the flow per hour; therefore, the units of $Q_1$ and $Q_2$ are gigajoules per hour (GJ / h).

The heat from the primary side ($Q_1$) was supposed to be equal to the heat from the main pipeline of the secondary side ($Q_2$), but there may exist some errors in the practical world. These errors could be attributed to the fact that water with a large flow rate cannot be fully heated and thus has low heat transfer efficiency. In contrast, if the flow rate is very small, the heat loss from the pipes to the air will increase, as the water with a smaller flow rate usually has a higher temperature and can easily lose heat. Moreover, excessively large flow rates are dangerous and may even lead to severe accidents because of the high pressure. Therefore, errors often exist during the heat transfer process, and the flow rates should be within a certain range.

Previously, researchers have reported various types of approaches to forecast the heat load of a system and control the heat source on its basis. In order to tune the flow rates using DRL, we introduced the target heat quantity $Q_{target}$ of an apartment according to the national standards\cite{China12,China19}: 

\begin{equation}
    Q_{target} = K \times S \times \frac{T_0-T_{out}}{T_0-\overline{T}} \times t
\end{equation}
where$\overline{T}$ is the designed outdoor temperature \renewcommand{\thefootnote}{\fnsymbol{footnote}}\footnote{$\overline{T}$ is defined as the daily average temperature over the last few decades with five non-guaranteed days per year\cite{China16}. For example, if we want to calculate $\overline{T}$ with the data of the last 30 years, firstly, we need to eliminate 150 of the coldest daily temperatures from the data, and then, we need to calculate the mean value of the daily temperatures of the rest of the days.}. $T_0$ is the required indoor temperature mandated by the government. $K$ is the index of the heat loss of the building, which implies the heat quantity required by the indoor heating equipment per unit area per unit time to ensure that the room temperature can achieve $T_0$, when the outdoor temperature is $\overline{T}$. $K$ depends on the geographical location and climate of a city. Its calculation principle involves thermal engineering\cite{China19}; therefore, we do not discuss it at length. $S$ is the heating area, $T_{out}$ is the current outdoor temperature, and t is the time. In this study, $K = 42.9 W/m^2, S = 6.2\times 10^4m^2, T_0 = 18^\circ C, t = 3600 s$, and $\overline{T}=-22.4^\circ C$. The values of $K$ and $\overline{T}$ are usually fixed for a city unless the government publishes new documents to change these values. The unit of $Q_{target}$ is also gigajoules per hour.

The system reaches its optimal state when $Q_{target} =Q_1=Q_2$ but is difficult to realize. For example, it is very difficult for operators to know how to tune the valves or the frequency of the pump correctly, because all the parameters in a heating system are interrelated. The occurrence of a change on one side may lead to other consequent changes in the entire system. For example, after turning up the valve opening on the primary side, the flow rate of this side increases. The heat transfer efficiency will decrease slightly because of the faster flow rate. Even though the supply water on the primary side as a whole carries more heat to the station, it is difficult to calculate the exact supply water temperature of the main pipes on the secondary side. In addition, the heat loss makes the DHS more complex. Consequently, simply establishing mathematical or statistical models for the heating process leads to significant errors. Therefore, we used deep neural networks (DNNs) to build the models. However, without a long period of adjustment, people still do not know what the optimal flow rates are even with these models. Therefore, deep reinforcement learning was used to comprehensively analyze these models and to provide optimization control immediately.

\section{District Heating System Model}
The parameters of a DHS, including the TD of both sides and the supply water temperature in the main pipeline on the secondary side, change when the flow rate of either side changes. We built models for them by using a deep neural network. It was essentially a regression process, and the mean square error (MSE) was the loss function. Historical data were divided into the training set and the testing set randomly. There were 16387 samples in the training set and 1955 samples in the testing set during the regression.  

\begin{figure}[htp]
	\begin{minipage}[t]{1\linewidth}
	\begin{center}
	 \includegraphics[scale = 0.2]{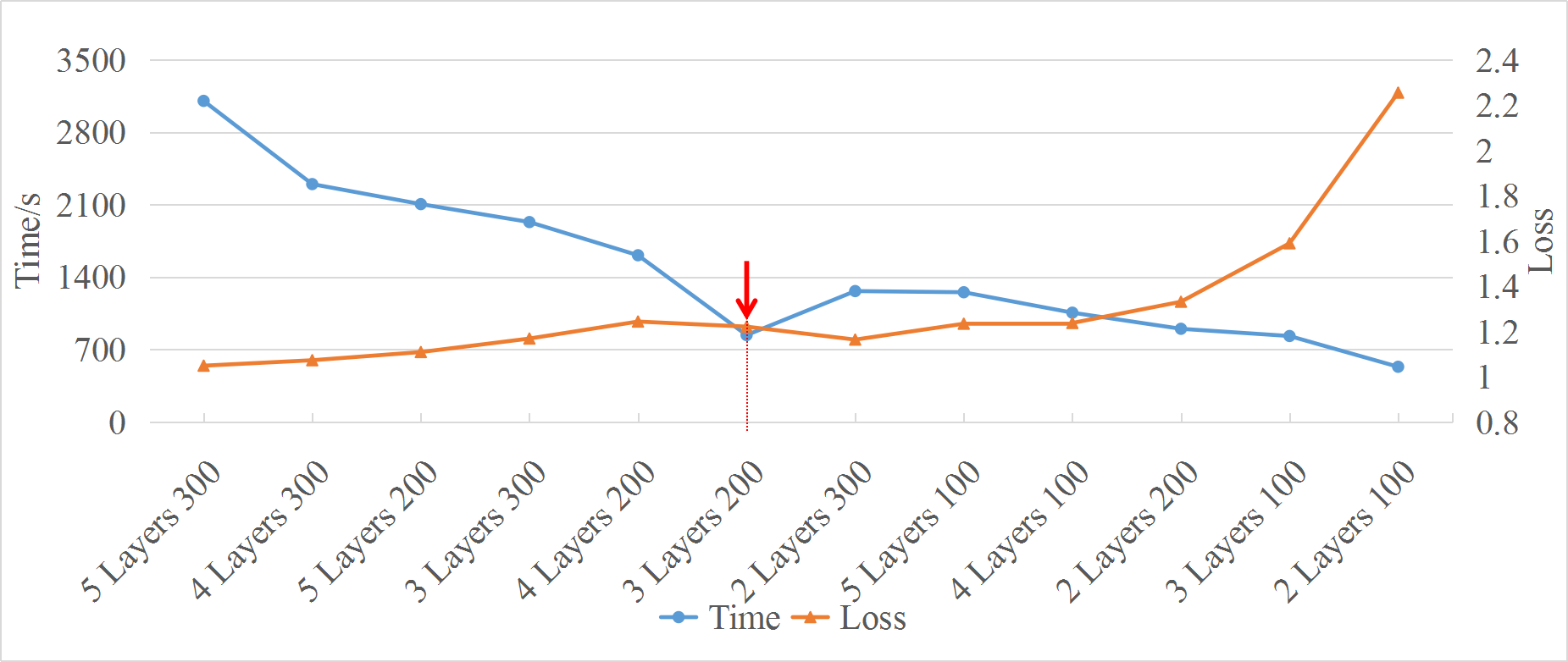}
	\end{center}
    \centerline{\scriptsize{(a) Supply Water Temperature on the Secondary Side (SWTS)}}   
	\end{minipage}

    \begin{minipage}[t]{1\linewidth}
		\begin{center}
  	  \includegraphics[scale = 0.2]{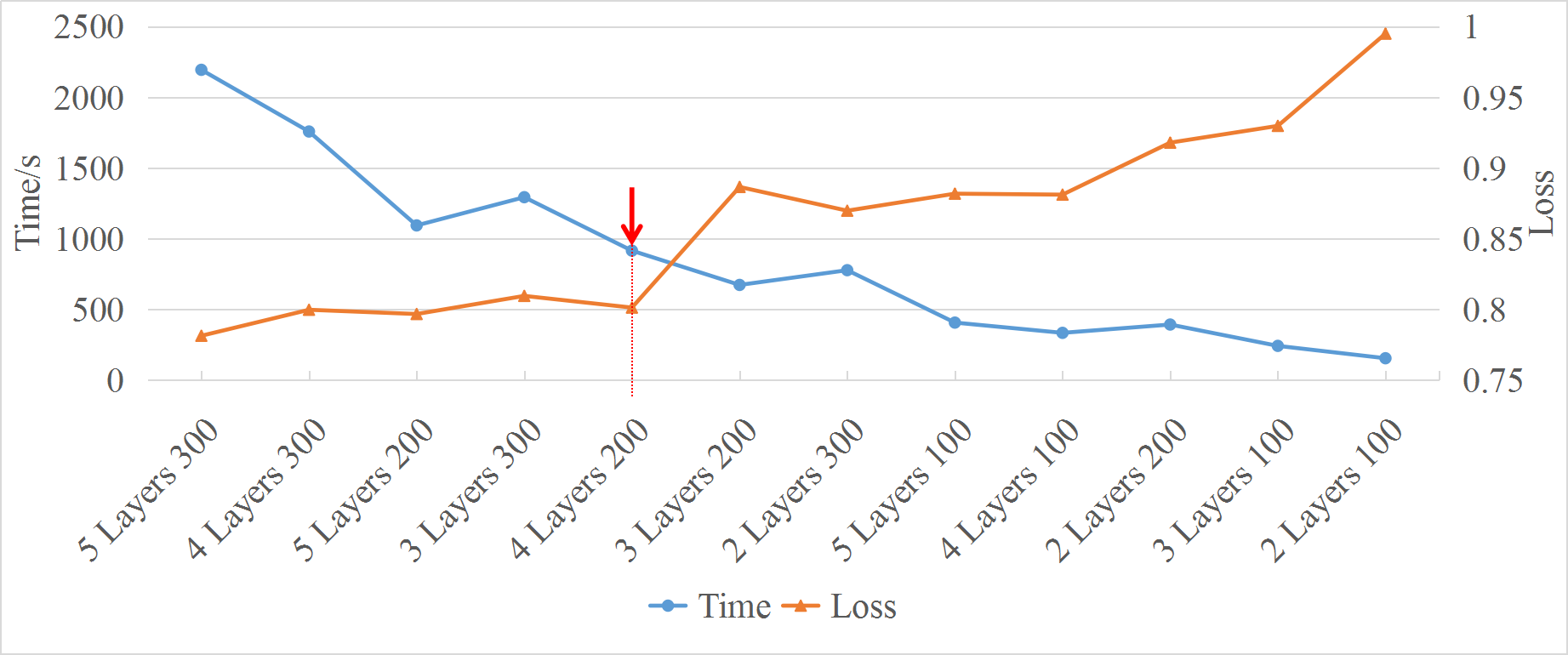}
  	  \centerline{\scriptsize{(b): Temperature Difference on the primary Side (TDP)}}
		\end{center}
	\end{minipage}

	\begin{minipage}[t]{1\linewidth}
		\begin{center}
  	 	\includegraphics[scale = 0.2]{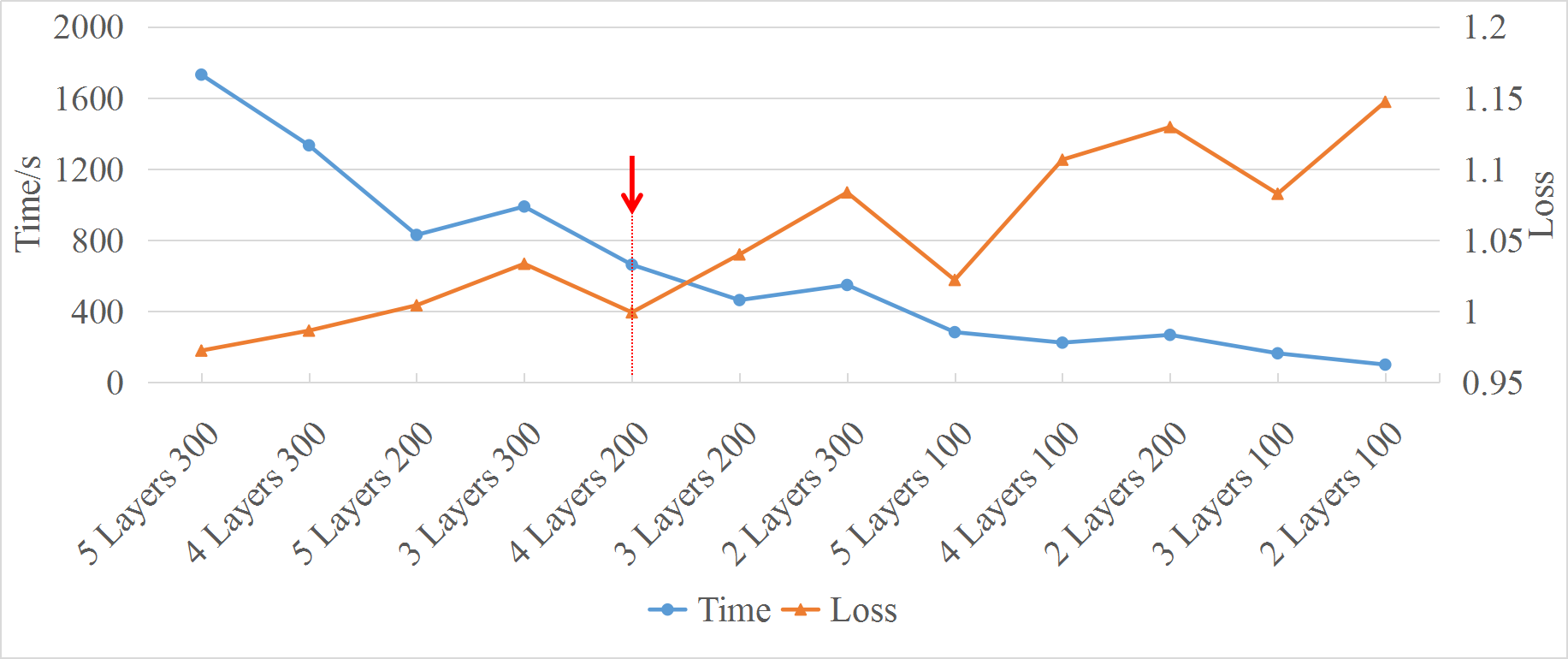}
  	 	\centerline{\scriptsize{(c): Temperature Difference on the Secondary Side (TDS)}} 
		\end{center}
	\end{minipage}
\caption{Loss Values and Time of networks with Different Structure}
\label{xuan}
\end{figure}

The modeling principles were based on the mechanism of DHS discussed in Section 2. For example, the input of the model for the TD on the primary side was the supply water temperature on the primary side and the flow rate on both the sides. The flow rate and the supply water temperature on the primary side reflected the heat quantity carried to the secondary side. Meanwhile, all of these three parameters influenced the heat transfer efficiency between the primary and the secondary side. Table~\ref{NNR} shows the input and output of the samples for these models, which were defined on the basis of the analysis of the DHS presented in Section 3. The MSE and the mean absolute error (MAE) are introduced as the evaluation indexes for each model.

It is generally known that a neural network with more nodes and layers tends to exhibit better performance but requires more training time. This is a trade-off problem between speed and accuracy; therefore, we have tried neural networks with different structures and selected the most suitable one for continued training. In particular, each network ran 2000 steps, and both the loss values and the time were considered in the evaluation. The networks that we finally used are shown in Figure~\ref{xuan} with red marks. The horizontal axis, i.e., "5 Layers 300," represents a network with five hidden layers and 300 nodes in each hidden layer.

We will need to tune the pump or the valve to control the flow rates indirectly in a real-world implementation. However, one heat source often matches more than one heat-exchange stations. If the valve on the primary side of the other stations changes, the flow rate on the primary side of our test station will also change even if the valve in our system remains unchanged. This is because the total amount of water from the heat source is kept nearly constant. Therefore, we tuned the valve opening by manual control and did not build a mathematical model for it. The flow rate on the secondary side was theoretically controlled by the frequency of the circulating pump; therefore, we estimated the flow with a polynomial fitting, as shown in equation~\ref{h}.
\begin{equation}\label{h}
    Flow_2 = 0.1492 \times f^2 -5.177 \times f +168.2
\end{equation}
However, some users like to open the tap of the heat-supply pipeline and collect free hot water for family use. This phenomenon along with pipeline damage leads to water leakage, and thus, the fitting performance is not very perfect. Operators may fine-tune the pump frequency after the DRL agent has taken the necessary action. 

\newcommand{\tabincell}[2]{\begin{tabular}{@{}#1@{}}#2\end{tabular}}
\begin{table*}[htb]\scriptsize 
\centering
\caption{Description and Evaluation about DHS Models}
\label{NNR}
\begin{tabular}{cccc}
\toprule  
Output & Input & MSE& MAE\\
\midrule 
Temperature Difference of the Primary Side (TDP)&\tabincell{c}{Flow rates on the both side\\Supply water temperature on the primary side} &0.489&0.503
\\
\midrule 
Temperature Difference of the Secondary Side (TDS)&\tabincell{c}{Supply Water Temperature on the secondary side\\the Flow of the Secondary Side\\Outdoor Temperature} &0.187&0.2457
\\
\midrule 
Supply Water Temperature on the secondary side (SWTS)&\tabincell{c}{Supply Water Temperature on the primary side\\ the Flow of the Primary and Secondary Sides }&0.6004&0.4864
\\
\midrule 
Flow on the secondary side&Polynomial Fitting &1.3495&0.8859
\\
\bottomrule 
\end{tabular}
\end{table*}

\section{Deep Reinforcement Learning}
In general, operators tune the hot water temperature from the heat source according to the outdoor temperature on the basis of their practical experience. The price of this rough heating control method leads to considerable waste of both water and heat energy. Nowadays, as people attach considerably more importance to the energy conservation issue, optimization control for the DHS has become very popular and requires simple and intelligent control methods. In this section, we will introduce the background knowledge on RL, how we use the deep deterministic policy gradient (DDPG) to deal with the heating control problems, and the environment we built for DDPG.

\subsection{Reinforcement Learning}
RL can solve sequential decision problems, which can often be modeled as Markov decision processes (MDPs)\cite{Bradtke94}. For example, at a time step $t$, an RL agent receives a state $s_t$ and then selects an action $a_t$ on the basis of $s_t$. The reward $r_t$ and the next state $s_{t+1}$ are obtained at the same time. Meanwhile, the cumulative discounted reward $R_t$ and the action-value function $Q^\pi(s_t,a_t)$ with the policy $\pi$ are defined as follows: 
\begin{equation}
  R_t=\Sigma^T_{i=t}{\gamma^{i-t}r(s_i,a_i)}
\end{equation}
\begin{equation}
  Q^\pi(s_t,a_t) = \mathbb{E}_\pi\left[ R_t|s_t,a_t \right]
\end{equation}
where $E$ denotes the expectation of the probabilities.
The agent continually makes actions until this episode is over. RL algorithms aim to obtain a policy $\pi$ that maximizes the expected cumulative discounted rewards from the initial position $R = \mathbb{E}\left[ \Sigma^T_{t=0}{\gamma^{t}r(s_t,a_t)} \right]$. 

When the policy for collecting the training data is the same policy network that is being learned, such an RL algorithm is called the on-policy. Otherwise, it is called the off-policy. Because of the higher training efficiency and better exploration, we preferred the off-policy RL in heating control problems. In addition, the action space in RL could be either discrete or continuous, and we adopted a continuous action to achieve more precise control. On the basis of the above mentioned demands, we used the deep deterministic policy gradient (DDPG) for our system.

\subsection{Deep Deterministic Policy Gradient}
DDPG\cite{Tuyen17} is an actor-critic algorithm and fully utilizes the advantages of the deterministic policy gradient (DPG)\cite{Silver14} and deep Q learning (DQN). An actor-critic algorithm includes an actor that performs an action on the basis of the current state and a critic that evaluates the action-value function $Q(s_t,a_t)$ according to the action performed by the actor. The actor in a traditional actor-critic algorithm is a stochastic policy $\pi(s_t|\theta^\pi)$, where $\theta^\pi$ is the parameter of the policy $\pi$. It generates a probability distribution of the action in the current state and chooses an action on the basis of this probability. 

In DPG, in contrast, the actor $\mu(s_t|\theta^\mu)$ is a deterministic policy and directly gives one certain action for each state instead of a probability distribution. However, for the given state and $\theta$, the trajectory generated by a deterministic policy remains unchanged, hence failing to explore the other trajectories. Therefore, additional noise is introduced for random exploration as shown in equation~\ref{c} and we choose the Ornstein-Uhlenbeck noise ($N_{OU}$) in our study. This policy $\beta$ is called the behavior policy, and its main function is to fully explore the environment and collect data for training the actor $\mu$. This is what we called the off-policy RL. 
\begin{equation}\label{c}
    \beta = \mu(s_t|\theta^\mu) + N_{OU}
\end{equation}
The DPG theorem gives the equation to update the actor network, as shown in equation~\ref{d}, while the update rule for the critic is shown in equation formulated on the basis of the Bellman equation~\ref{e}:

\begin{equation}\label{d}
    \nabla_{\theta^\mu}J_\beta(\mu) \approx \mathbb{E}_{s_t\sim\rho^\beta}\left[ \nabla_a Q(s_t,a|\theta^Q)|_{a=\mu(s_t)}\nabla_{\theta_\mu}\mu(s_t|\theta^\mu) \right]
\end{equation}
\begin{equation}\label{e}
    L = \mathbb{E}_{s_t\sim\rho^\beta,a_t\sim\beta}\left[ (y_t-Q(s_t,a_t|\theta^Q))^2 \right]
\end{equation}
where $y_t = r(s_t,a_t)+\gamma Q(s_{t+1},\mu(s_{t+1})|\theta^Q)$, $\rho^\beta$ is the probability distribution function of the state based on the behavior policy $\beta$.

DDPG combines the two techniques of DQN with DPG to improve its performance: experience replay and target networks. Experience replay can break the temporal correlations among the collected data, and the target networks can help the algorithms become more stable and achieve faster convergence. For each mini-batch, the policy gradient of the actor network and the loss that the critic network aims to minimize are shown in equation~\ref{f} and equation~\ref{g}, respectively:

\begin{equation}\label{f}
   \nabla_{\theta^\mu}J_\beta(\mu) \approx \frac{1}{N}\mathop{\Sigma}\limits_i(\nabla_a Q(s_i,a|\theta^Q)|_{a=\mu(s_i)}\nabla_{\theta_\mu}\mu(s_i|\theta^\mu) )
\end{equation}
\begin{equation}\label{g}
    L = \frac{1}{N}\mathop{\Sigma}\limits_i((y_i-Q(s_i,a_i|\theta^Q))^2)
\end{equation}
where $y_i=r_i+\gamma Q'(s_{i+1},\mu'(s_{i+1}|\theta^{\mu'})|\theta^{Q'})$, $\mu'$ and $Q'$ belong to the target networks.

The target actor networks and the target critic networks update as follows:
\begin{equation}
    \theta^{\mu'}\leftarrow \tau\theta^\mu+(1-\tau)\theta^{\mu'}
\end{equation}
\begin{equation}
    \theta^{Q'}\leftarrow \tau\theta^Q+(1-\tau)\theta^{Q'}
\end{equation}
where $\tau$ is an update parameter that is usually considerably less than 1.

In fact, there are some other DRL algorithms for continuous control. In this study, we applied the proximal policy optimization (PPO2)\cite{Schulman17} and soft actor-critic (SAC)\cite{Haarnoja18} to control the system and compared its performance with DDPG's.

\begin{figure*}[htb]
  \centering
  \includegraphics[width=14cm]{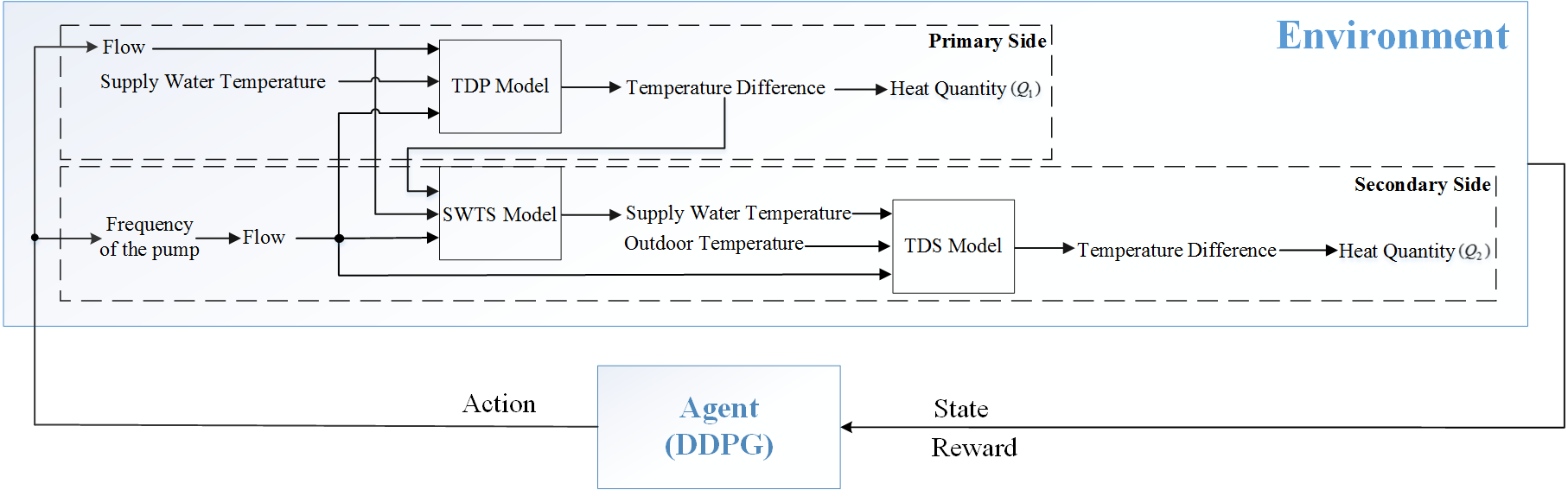}
\caption{Deep Reinforcement Learning based on Heating Control}
\label{fig:FC}
\end{figure*}

\subsection{Customized Environment for DDPG}
We built a customized environment based on OpenAI Gym\cite{Brockman16} for the implementation of DDPG. A reinforcement learning environment contains three major parts: state, action, and reward.
\begin{itemize}
    \item \textbf{State}: The state includes the outdoor temperature, supply water temperature on the primary side, and the target heat quantity. 
    \item \textbf{Action}: It is a two-dimensional array, and the range of both the elements is -1 to 1. The first dimension controls the flow rate of the primary side $Flow_1$ by equation~\ref{h}, and the second dimension controls the frequency of the circulating pump $f$ by equation~\ref{i}. The maximum and minimum of these two variables are decided on the basis of the practical results. For example, the maximum of the pump frequency was 43.88 in our historical data obtained by manual control. However, sometimes, the frequency exceeded 43.88 in order to reach its optimal state. Therefore, we expanded the range for $Flow_1$ and $f$.
    \begin{equation}\label{h}
        Flow_1 = 55 + 45 \times action[0]
    \end{equation}
    \begin{equation}\label{i}
        f = 35 + 15 \times action[1]
    \end{equation}
    \item \textbf{Reward}: We defined three kinds of reward functions as follows:
    \begin{equation}\label{j}
           R_{Q_1} = -(\vert Q_1 - Q_{target} \vert ).
    \end{equation}
    \begin{equation}\label{kkk}
           R_{Q_2} = -(\vert Q_2 - Q_{target} \vert ). 
    \end{equation}
    \begin{equation}\label{l}
    \begin{split}
           R_{Q_1Q_2} = -\frac{(\vert Q_1 - Q_{target} \vert + \vert Q_2 - Q_{target}\vert)}{2}.  
    \end{split}
    \end{equation}
    Equation~\ref{j} and equation~\ref{kkk} express the agent that controls the flow rates according to the heat quantity of only one of the two sides, while the agent with the reward function~\ref{l} tunes the flow rates on the basis of the heat quantity of both the sides. We conducted a further comparison of the total rewards with different reward functions. The total reward was the accumulated reward value of an episode. In this study, each episode contained 500 samples; therefore, the total reward was the sum of the reward values of 500 samples.
\end{itemize}
The DDPG control flow chart is shown in Figure~\ref{fig:FC}, where the numbers suggest the computational order in the code.

\section{Implementation}
\begin{figure}[htb]
  \centering
  \includegraphics[width=8cm]{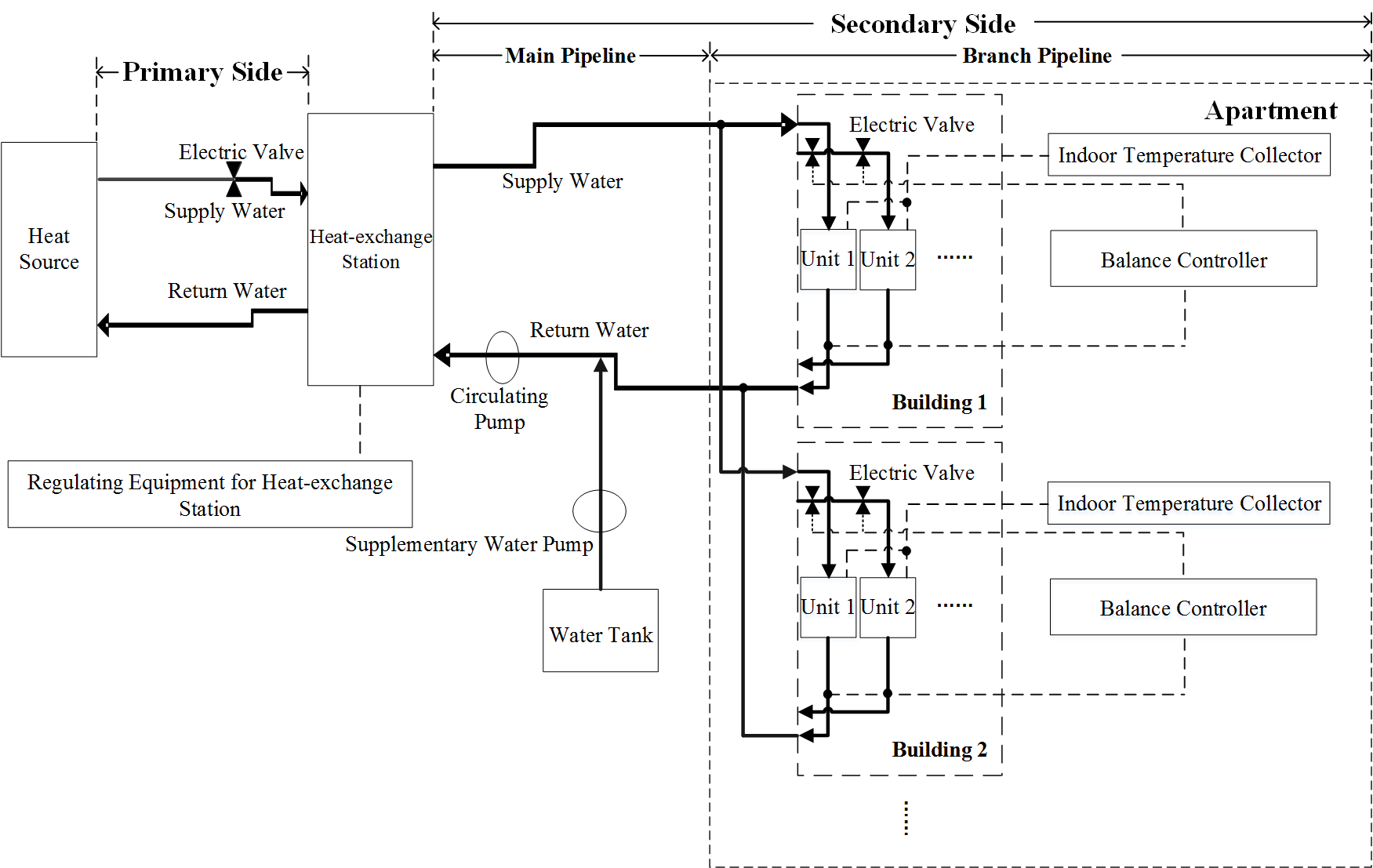}
\caption{Heat Control System Based on NB-IoT}
\label{fig:heat_exchange}
\end{figure}

We designed a control system for the DHS. The structure of this heating control system is shown in Figure~\ref{fig:heat_exchange}. It contains three major parts: balance controllers installed in the apartment, regulating equipment in the heat-exchange station, and indoor temperature sensors.

\subsection{NB-IoT and Alibaba Cloud}
We used NB-IoT as the wireless communication approach because of its low cost and ease of deployment. The heating control system collects and sends heating parameters, including flow rates and water temperature, to the cloud for people to monitor. The related data transmitted by NB-IoT are stored in Alibaba Cloud. These parameters are very important for us to further improve the system, e.g., for training DRL algorithms. Moreover, Alibaba Cloud with NB-IoT allows operators to monitor the system situation and control it from their office. 

Alibaba Cloud has been widely used in the marketplace and has many success stories with famous companies such as Ford, ESRI, and PicsArt. Alibaba Cloud Database services offer customers data backup, recovery, monitoring, migration, and disaster recovery solutions, thereby ensuring good reliability and stability. Furthermore, Alibaba Cloud adheres to international information security standards to ensure a high level of security compliance. From the aspect of customer privacy, Alibaba Cloud was GDPR ready by the effective date of May 25, 2018 and is committed to the protection of personal information. Because of the above-mentioned advantages, we used Alibaba Cloud as the cloud platform in our system. Table~\ref{data} shows the parameters that we stored on Alibaba Cloud.

To sum up, Alibaba Cloud connected by NB-IoT allows operators to remote monitor the DHS. At the same time, the data stored on the cloud is of great value for us to understand the heating system. It also allows operators to fine tune the system if necessary. In addition, the entire regulation process saves more human resources.

\begin{table*}[htb]\footnotesize 
\centering
\caption{Available Data Stored in Alibaba Cloud}
\label{data}
\begin{tabular}{ccc}
\toprule  
\multirow{6}*{Primary Side}&\multicolumn{2}{c}{Supply Water Temperature}\\
                           &\multicolumn{2}{c}{Return Water Temperature}\\
                           &\multicolumn{2}{c}{Temperature Difference}\\
                           &\multicolumn{2}{c}{Flow}\\
                           &\multicolumn{2}{c}{Heat Quantity}\\
                           &\multicolumn{2}{c}{Valve Opening}\\
\midrule
\multirow{9}*{Secondary Side}&\multirow{6}*{Main Pipeline}&Supply Water Temperature\\
                             &                            &Return Water Temperature\\
                             &                            &Temperature Difference\\
                             &                            &Flow\\
                             &                            &Heat Quantity\\
                             &                            &Frequency of Circulating Pump\\  \cline{2-3}
          &\multirow{3}*{Branch Pipeline (for each Unit)}&Valve Opening\\
          &                                               &Return Water Temperature\\
          &                                               &Indoor Temperature (Only Some Users)\\
\midrule 
\multirow{2}*{Shared Data}& \multicolumn{2}{c}{Outdoor Temperature}\\
						   & \multicolumn{2}{c}{Target Heat Quantity}\\	
\bottomrule 
\end{tabular}
\end{table*}

\subsection{Balance Controllers}
\begin{figure}[htb]
  \centering
  \includegraphics[width=8cm]{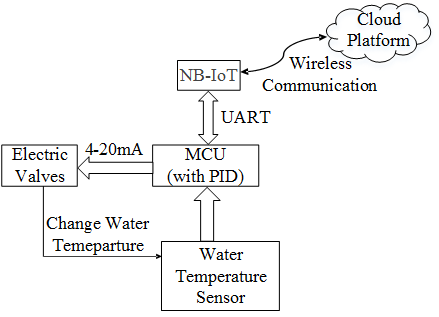}
\caption{Balance Controller for Secondary Side}
\label{fig:Balance_C}
\end{figure}

The balance controllers installed in the apartment can send the return water temperature and valves opening to Alibaba Cloud through NB-IoT and, at the same time, control the electric valves by the PID algorithm. A balance controller contains an MCU, an electric valve, a water temperature sensor, and an NB-IoT module, which is shown in ~\ref{fig:Balance_C}.

MCU contains the PID algorithm that can control the electric valve opening. Electric valves are installed on the supply water side in the branch pipeline, while water temperature sensors are installed on the return water side. The valve opening is controlled by the current going through it. It will be in a completely open or close state when the current is 20 mA and 4 mA, respectively. Balance controllers can make the return water temperature of different units the same and finally realize the hydraulic balance for an apartment complex.

\subsection{Regulating Equipment for Heat-exchange Station}
\begin{figure}[htb]
  \centering
  \includegraphics[scale = 0.75]{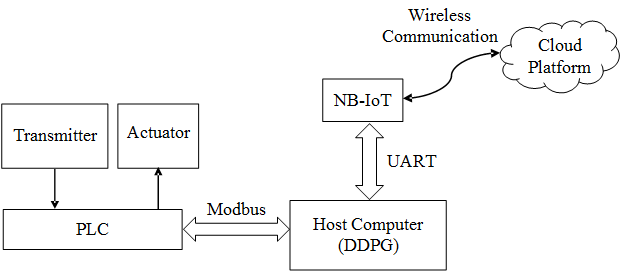}
\caption{Regulating Equipment for Heat-exchange Station}
\label{fig:Regualt_E}
\end{figure}

The regulating equipment controls the system as a whole, and the details are shown in Figure~\ref{fig:Regualt_E}. There are many transmitters in the heat-exchange station to measure the heating parameters of the main pipeline. These transmitters are connected with PLC through the I/O interface and can send these parameters to PLC\cite{Zhang18}. The regulating equipment sends the heating parameters to the cloud. Moreover, PLC can generate electric valve signals to control the heat-exchange station. PLC communicates with the host computer with the Modbus protocol. Therefore, the operators can manually control the valves installed both in the apartment and on the primary side or the circulating pump through the host computer. We intend to use the host computer to provide automatic control by DDPG in the future.

\subsection{Indoor Temperature Sensor}

\begin{figure}[htb]
  \centering
  \includegraphics[scale = 0.7]{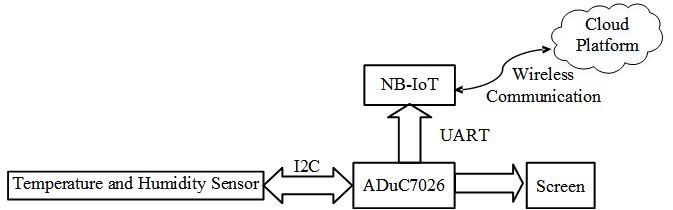}
\caption{Indoor Temperature Collector}
\label{fig:IDC}
\end{figure}

Indoor temperature sensors, as shown in Figure~\ref{fig:IDC}, are installed in houses and measure the indoor temperature and humidity at regular intervals. They contain a micro control unit (MCU) sensor, sensors, a screen, and an NB-IoT wireless communication module. To begin with, the data are read by the MCU through I2C. The data will be processed, analyzed, and transmitted to the cloud through the NB-IoT module. The MCU communicates with the NB-IoT through UART. 

\section{Evaluation}
We conducted a real-world case study in a city at latitude 46$^{\circ}$ by using the proposed heating control system. The total heating area of the test station was approximately 62,000 $m^2$ with 10 buildings, and each building had several units. Other details about the test station are shown in Table~\ref{buhaoshi}. We tuned the flow rates manually at that time, and recently, we used the data to train and test the DDPG in the simulation experiments. 

\begin{table}[htb]\scriptsize
\centering
\caption{Details about Test Station}
\label{buhaoshi}
\begin{tabular}{cccc}
\toprule  
Name&Type&\tabincell{c}{the Number\\ of Units}&\tabincell{c}{Building\\ Area}\\
\midrule 
Building 3&Residential Building&5&4849
\\
Building 4&Residential Building&4&3871
\\
Building 5&Residential Building&3&2948
\\
Building 6&Residential Building&4&4271
\\
Building 7&Residential Building&9&8672
\\
Building 8&Residential Building&4&4283
\\
Building 9&Residential Building&4&4337
\\
Building 16&Residential Building&6&7568
\\
Building 19&Residential Building&5&3567
\\
X Company&Public Office&*&17834
\\
\bottomrule 
\end{tabular}
\end{table}

\subsection{DDPG Training}
The dataset obtained after the manual control was from January 15 to April 20, in 2018. Each sample for training or testing the DDPG included the supply water temperature on the primary side, outdoor temperature, and, the target heat quantity. We divided the samples into the training set and the testing set. In particular, the samples of the first seven days were put into the training set, and the sample of the eighth day was assigned to the testing set, and so on. As a result, the training set included the data of 84 days, and those of the other 12 days in the testing set. Each day had more than 24 samples because the sampling interval in the case study was less than 1 h. The training set had 16387 samples, while the testing set had 1955 samples. 

\begin{figure}[htb]
\centering
\subfigure[]{
\begin{minipage}[t]{0.5\linewidth}
\centering
\includegraphics[scale = 0.12]{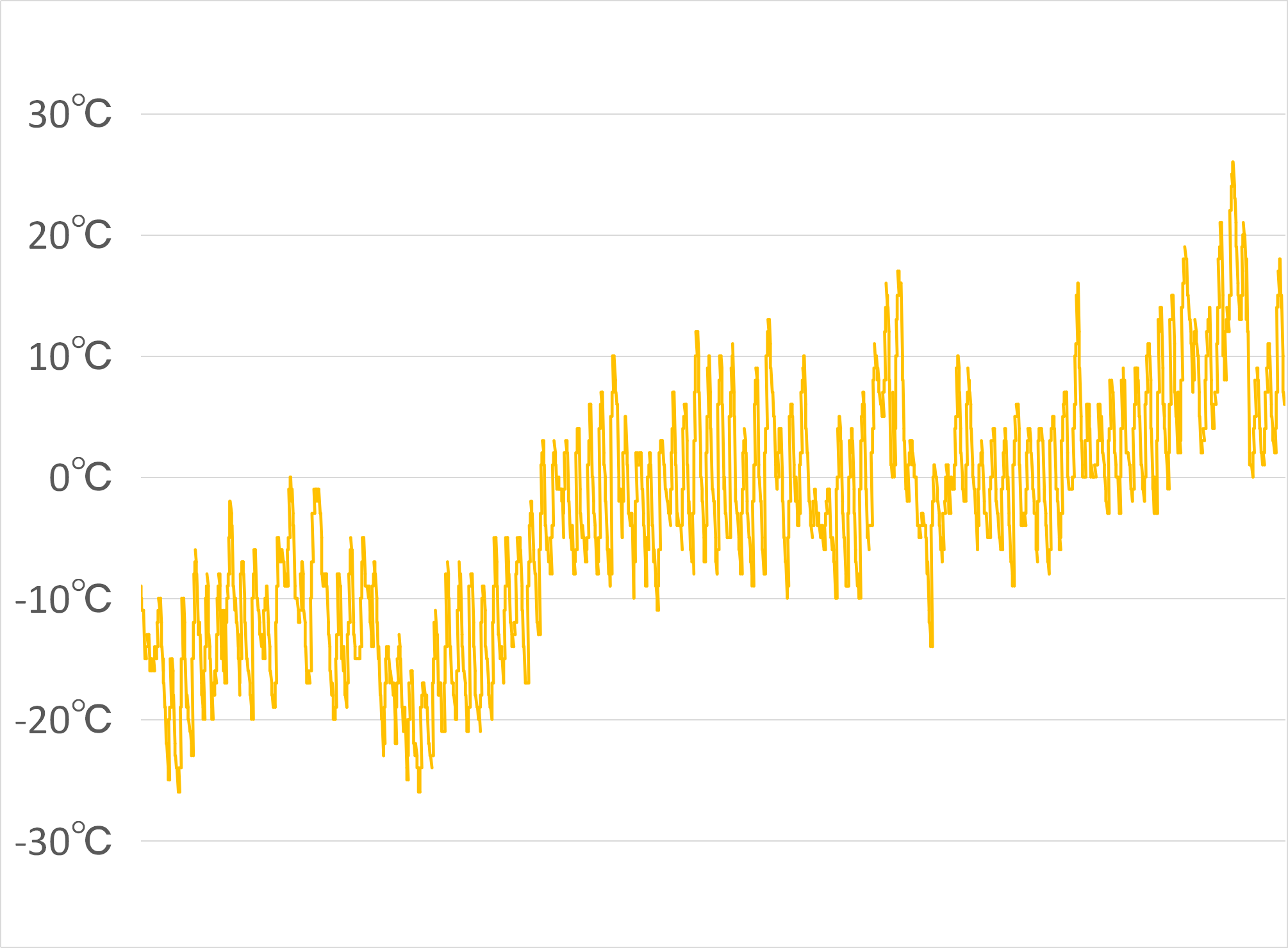}
\end{minipage}%
}%
\subfigure[]{
\begin{minipage}[t]{0.5\linewidth}
\centering
\includegraphics[scale = 0.12]{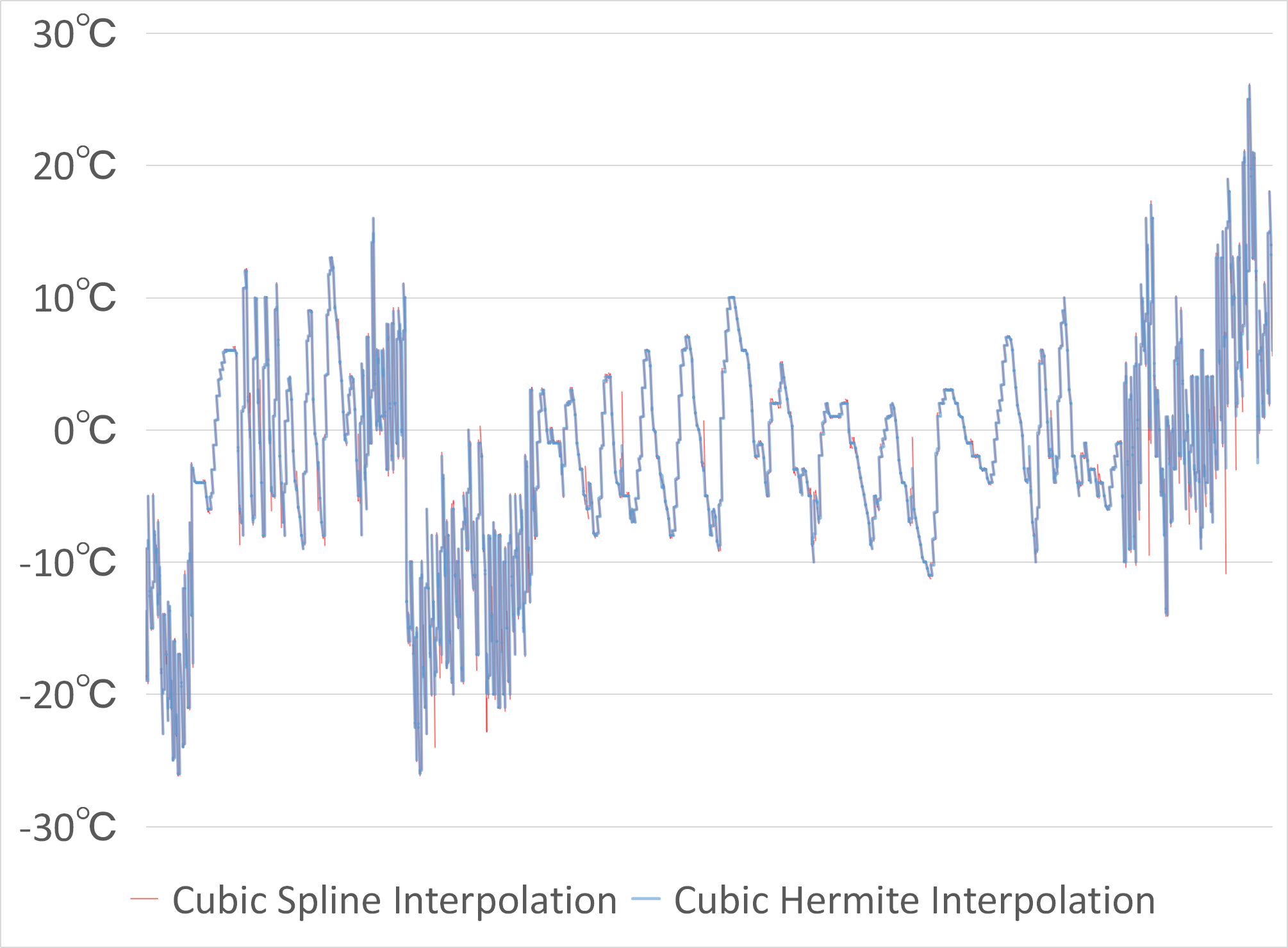}
\end{minipage}%
}%
\caption{Existing Data and Interpolation Results: (a) Existing Outdoor Temperature; (b) Interpolation Results.}
\label{chaz}
\end{figure}

One problem was the lack of sufficient outdoor temperature data. The outdoor temperature was recorded in the cloud nearly once every hour. We only had 2364 samples with their corresponding outdoor temperature. The total number of the samples was 18342, and most samples did not have the outdoor temperature. To solve this problem, we used the piecewise cubic Hermite interpolation and the cubic spline interpolation to calculate the outdoor temperature for every sample, as shown in Figure~\ref{chaz}. There were many spikes after cubic spline interpolation, but the outdoor temperature was supposed to change without so many sudden changes. Therefore, we used the result of piecewise cubic Hermite interpolation.

\begin{figure}[htb]
	\begin{minipage}[t]{1\linewidth}
	\begin{center}
	 \includegraphics[scale = 0.3]{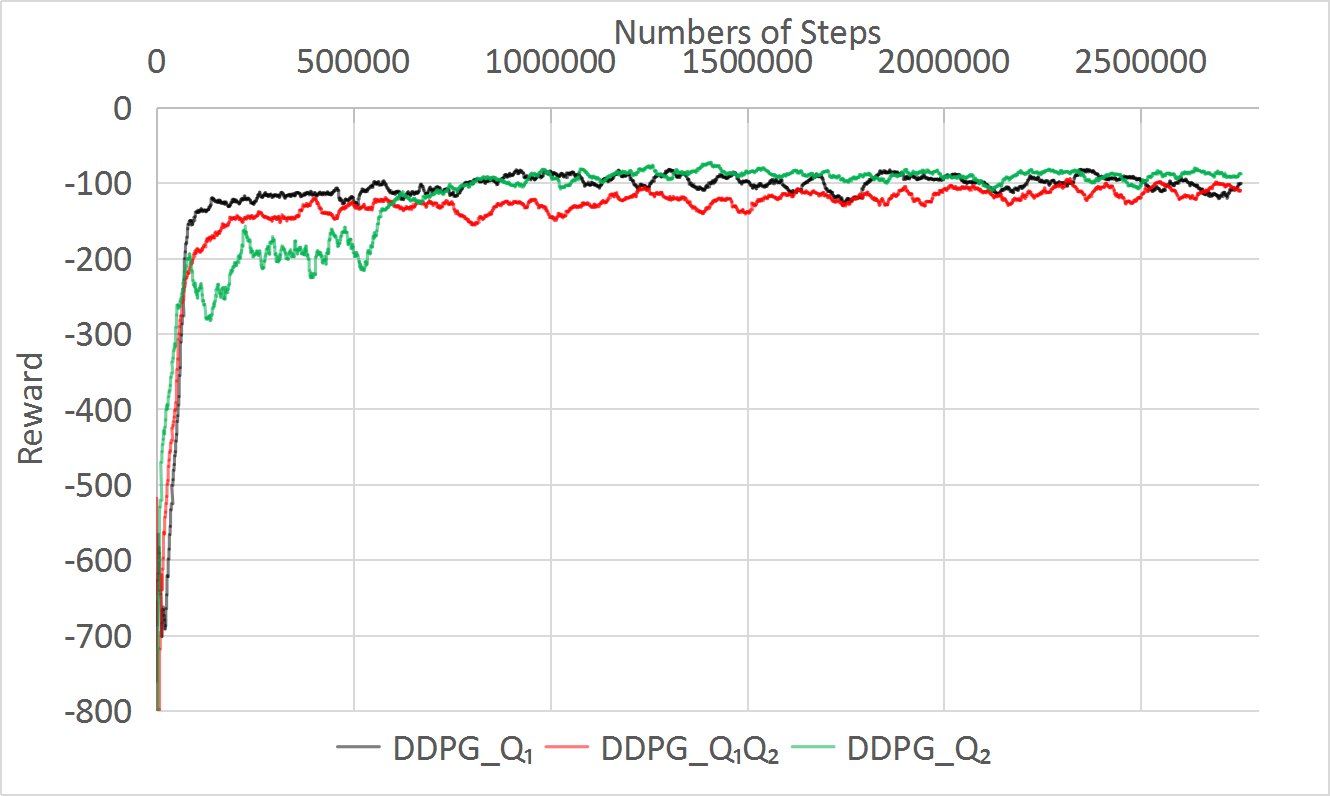}
	\end{center}
    \centerline{\scriptsize{(a) Learning Curve}}   
	\end{minipage}

    \begin{minipage}[t]{1\linewidth}
		\begin{center}
  	  \includegraphics[scale = 0.16]{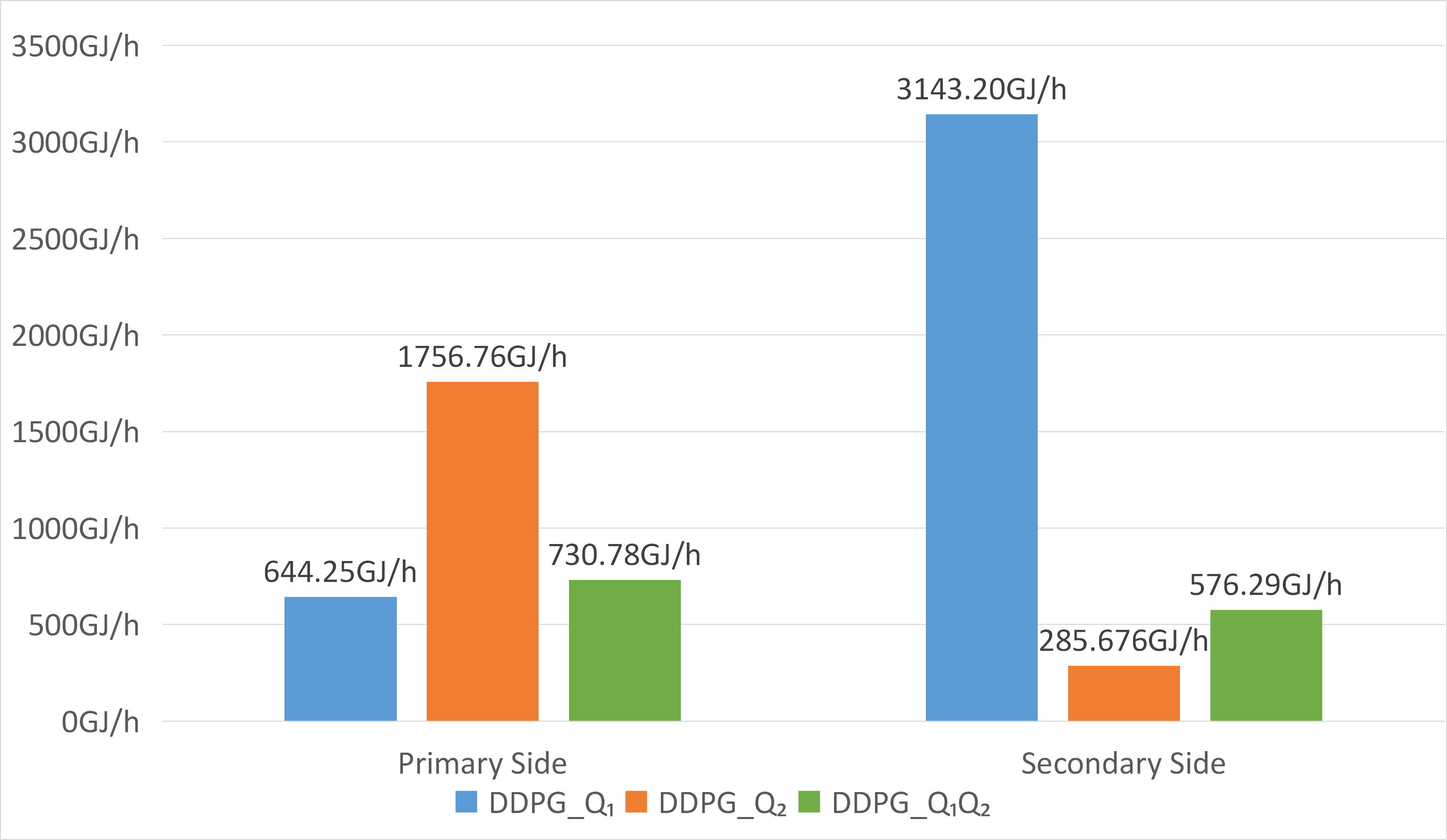}
  	  \centerline{\scriptsize{(b): Cumulative Error}}
		\end{center}
	\end{minipage}
\caption{Comparison Between Different Reward Functions}
\label{reward}
\end{figure}

Before conducting further tests, we trained the DDPG with three different reward functions, as in Section 4, to find the optimal one. The cumulative error (CE) was defined in equation~\ref{k} to evaluate the performance of the agents. 
\begin{equation}\label{k}
           CE= \Sigma^N_{i}{\vert Q^i-Q^i_{target}\vert},
\end{equation}
where N denotes the number of the samples and N = 1954 in this study. The lower the CE was, the smaller was the difference between the target and the real heat quantity in the testing set, thus indicating a better performance for the agent. Figure~\ref{reward} shows the learning curves of three agents and their CE values. Only the agent with a reward function~\ref{l} could make the heat quantity of both the sides approach the target heat quantity. Therefore, we applied the reward function~\ref{l} in this study.

\subsection{Other Algorithms for Comparison}
We conducted the flow rate control with other algorithms to show the superiority of the performance of DDPG in heating control problems. For deep reinforcement learning, we chose PPO2 and SAC for comparison, as both of them are advanced algorithms for continuous control. Their training methods are the same as those for DDPG.

\begin{figure}[htb]
  \centering
  \includegraphics[scale = 0.25]{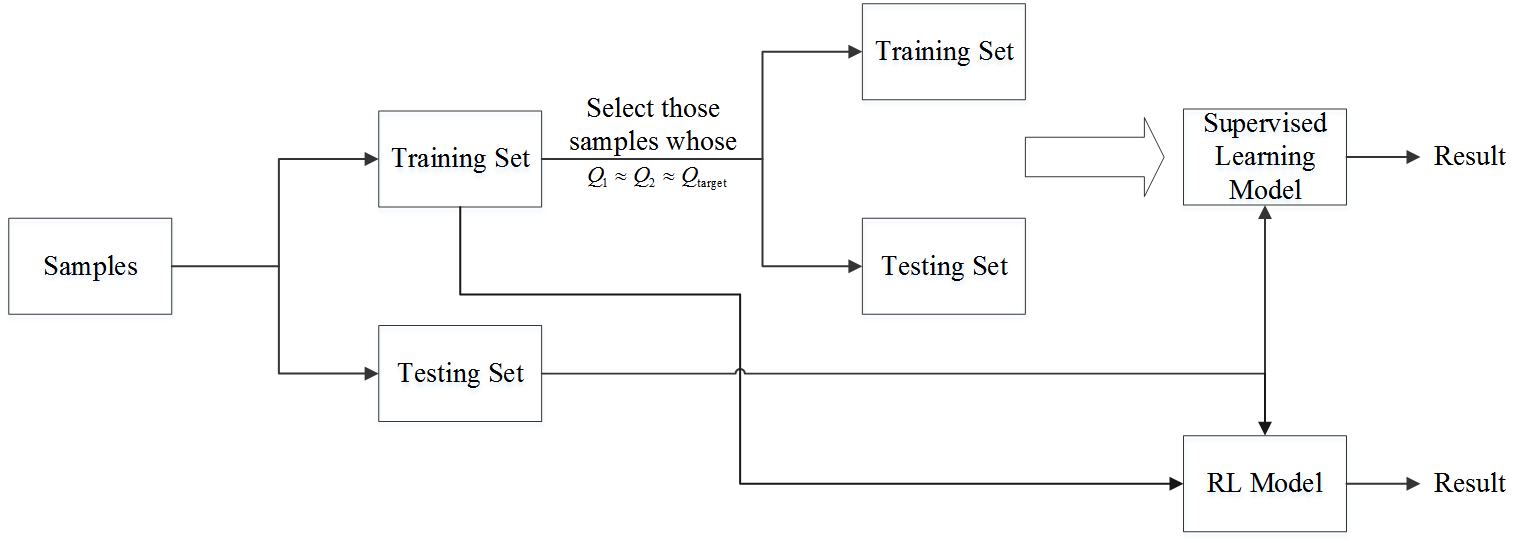}
\caption{Evaluation for Different Models} 
\label{diff}
\end{figure}

Moreover, we used a support vector machine (SVM) and DNN as the supervised learning (SL) methods for heating control. More concretely, we selected the samples whose $Q_1$ and $Q_2$ were close to their $Q_{target}$ from the training set. Obviously, the selected samples approached the optimal states. For the primary side, the input data were the outdoor temperature, target heat quantity, and the supply water temperature of the primary side, while the output was its flow. Similarly, the input data for the secondary side contained the outdoor temperature, target heat quantity, and the supply water temperature of the secondary side and the flow rate of the primary side. We took the same testing set, which was used to evaluate the performance of DDPG, to assess these supervised learning models. Figure~\ref{diff} intuitively presents the method used to train and evaluate different models in our study. It can be easily seen that DRL could make full use of the data because it did not require "labels" during learning. 

\subsection{Results}
\begin{figure}[htb]
  \centering
  \includegraphics[scale = 0.25]{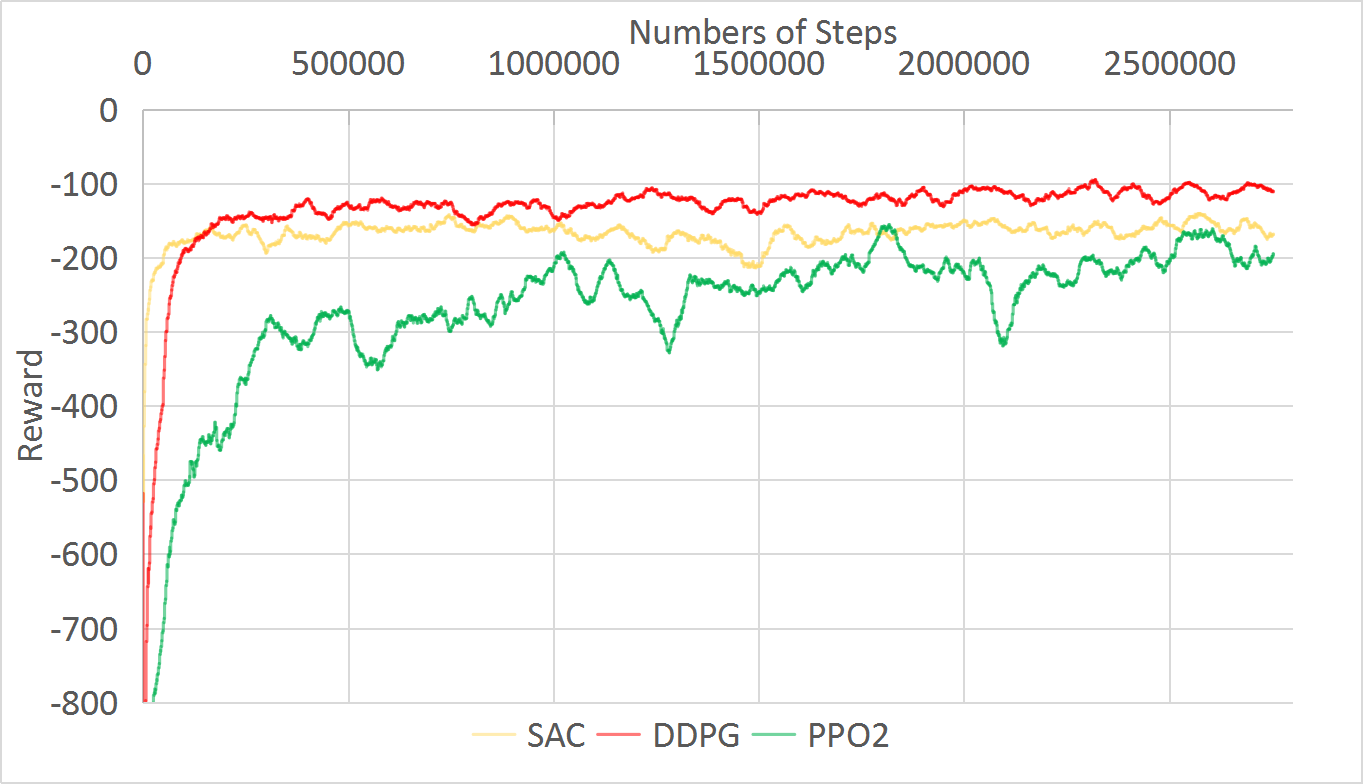}
\caption{Total Rewards of Training Set}
\label{fig:reward}
\end{figure}

The total rewards of DDPG, PPO2, and SAC are shown in Figure~\ref{fig:reward}. A further comparison of the testing set included two aspects: water usage and heat energy. We also provided some ideas about DDPG training in a real-world implementation.

\begin{figure}[htb]
  \centering
  \includegraphics[width=8.5cm]{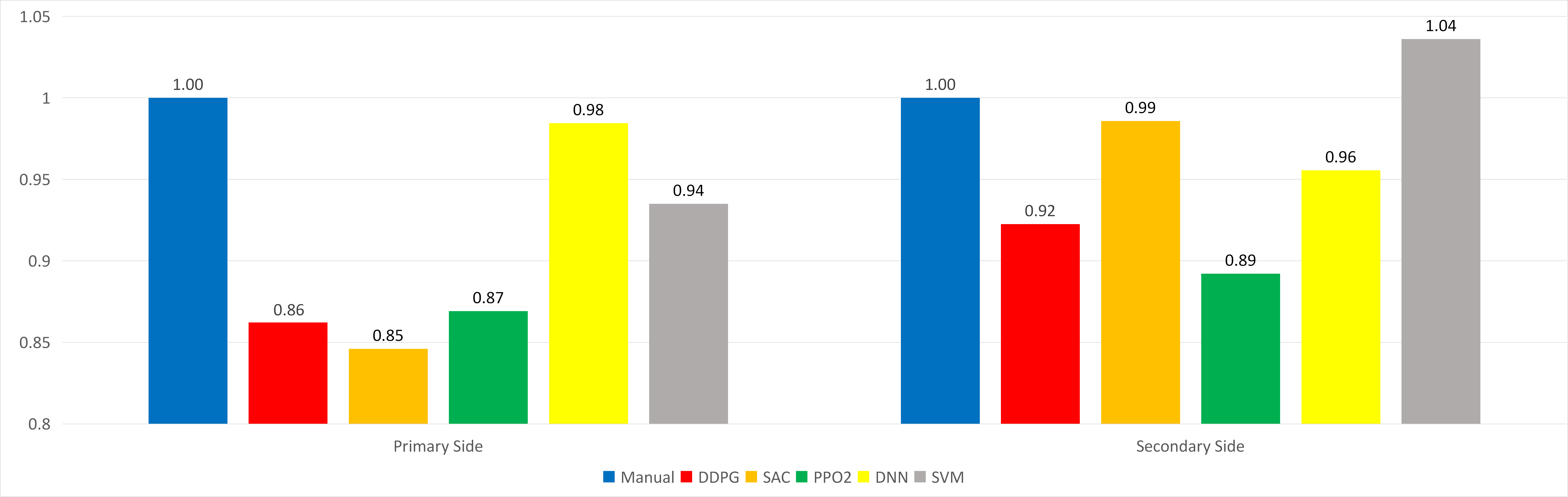}
\caption{Normalized Total Water Consumption} 
\label{zongshui}
\end{figure}

\textbf{Water Usage Comparison.} We calculated the normalized total water consumption by humans and by algorithms in order to present more intuitive results. Specifically, each result was divided by the water consumption of manual control. The result was shown in Figure~\ref{zongshui}. The water consumption controlled by humans was 127233.38 t/h on the primary side and 319806.51 t/h on the secondary side. Compared with manual control, water consumption with DDPG control was by 86\% of it on the primary side and 92\% of it on the secondary side. The Water consumption of DDPG was 404763.43 t/h, and it meant DDPG can save 42276.45 tons of water per hour. The water consumption using DDPG control was the second least among other algorithms.

\begin{figure}[htb]
\centering
\subfigure[]{
\begin{minipage}[t]{0.5\linewidth}
\centering
\includegraphics[width=4.0cm]{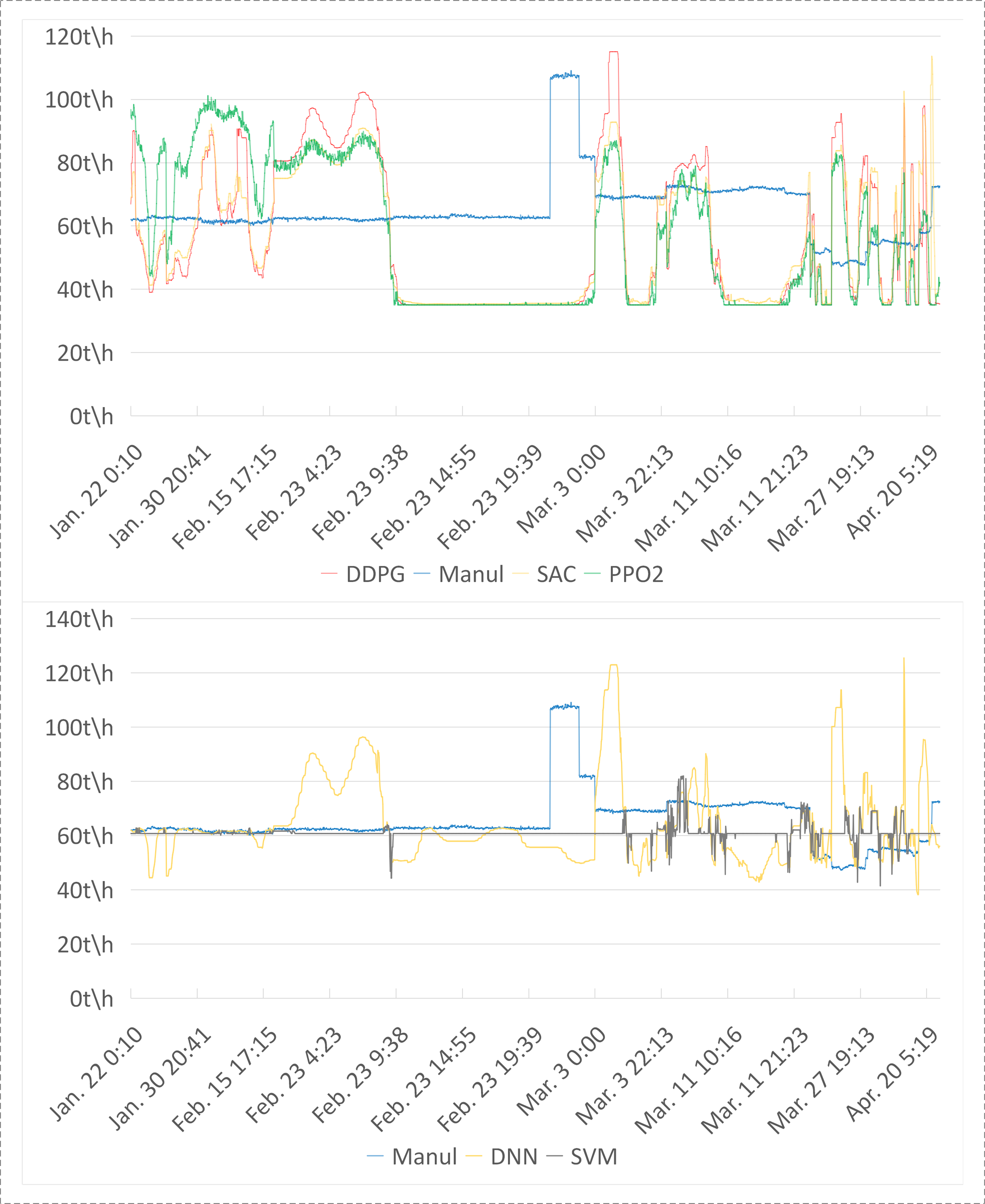}
\end{minipage}%
}%
\subfigure[]{
\begin{minipage}[t]{0.5\linewidth}
\centering
\includegraphics[width=4.0cm]{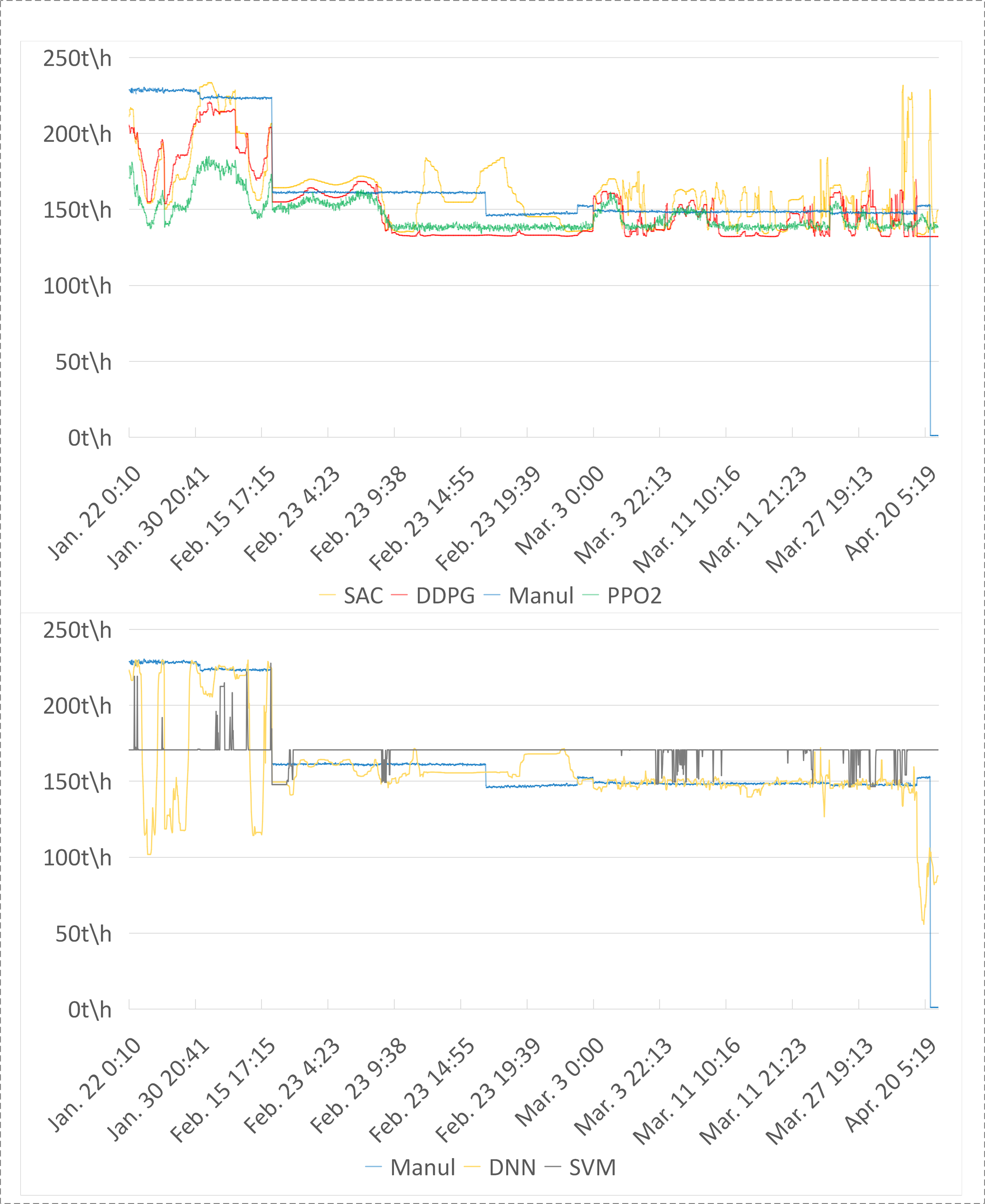}
\end{minipage}%
}%
\caption{ Daily Water Consumption: (a) Primary Side; (b) Secondary Side.}
\label{fig:res}
\end{figure}

Furthermore, the daily water consumption per hour is shown in Figure~\ref{fig:res}. Manual control did not change the flow rates frequently, hence leading to considerable water wastage. In contrast, DDPG as well as the other algorithms tuned the flow rates according to the outdoor temperature and the supply water temperature on the primary side in different ways. 

\begin{figure}[htb]
\centering
\subfigure[]{
\begin{minipage}[t]{0.5\linewidth}
\centering
\includegraphics[width=3.9cm]{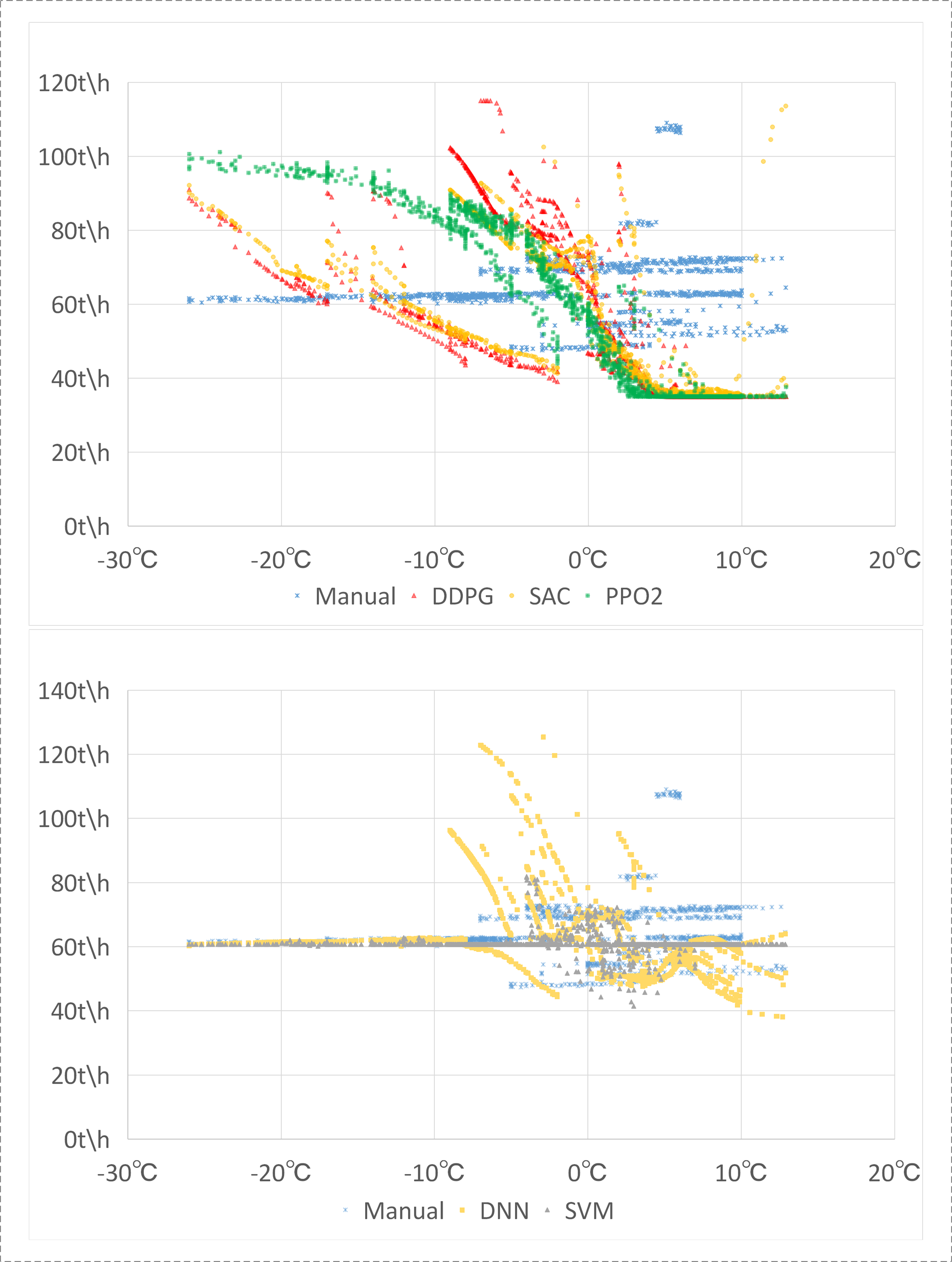}
\end{minipage}%
}%
\subfigure[]{
\begin{minipage}[t]{0.5\linewidth}
\centering
\includegraphics[width=3.9cm]{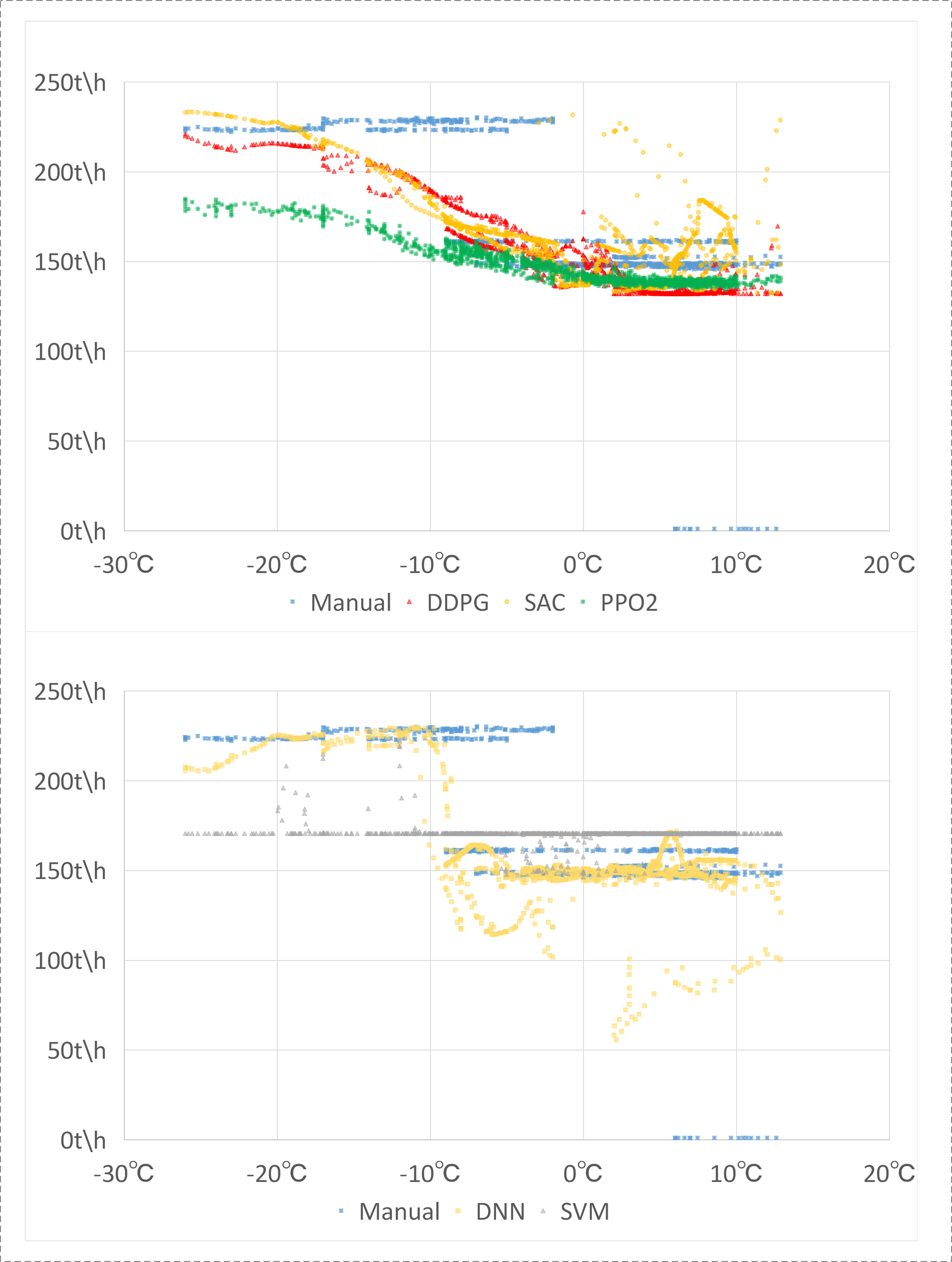}
\end{minipage}%
}%
\caption{ Flow rate-Temperature Scatter Plot: (a) Primary Side; (b) Secondary Side.}
\label{fig:sandian}
\end{figure}

In order to further explain which algorithms exhibited the best performance, the flow rate-temperature scatter diagrams are shown in Figure~\ref{fig:sandian}. This figure shows that DRL could regulate the flow rates appropriately compared with the manual control, as the flow rates decreased when the temperature increased. Note that there was a sudden change for DRL on the primary side when the temperature ranged from -10$^{\circ}$C to 0$^{\circ}$C. This was attributed to the fact that the supply water temperature from the boiler house decreased, and thus, the flow rate on the primary side increased to bring sufficient heat energy to the secondary side. In contrast, the flow rates under SL control failed to be tuned reasonably on the basis of the outdoor temperature. 

\begin{figure}[htb]
\centering
\includegraphics[width=8.5cm]{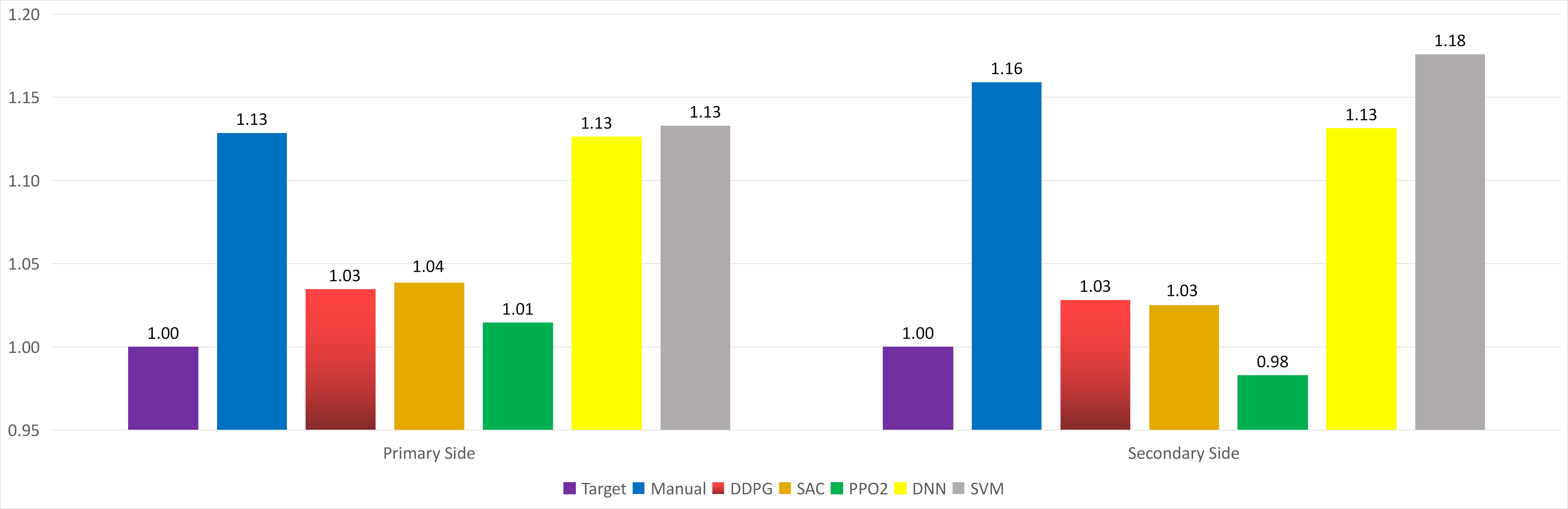}
\caption{Normalized Total Heat Consumption} 
\label{zongre}
\end{figure}

\textbf{Heat Energy Consumption.} Similarly, the total heat energy consumption controlled by different methods was divided by the target heat quantity, and the normalized heat energy consumption is presented in Figure~\ref{zongre}. Compared with the manual control, the heat quantity decreased by 9.7\% on the primary side and 12.6\% on the secondary side. The total heat energy consumption of DDPG was 18219.17 GJ/h and was the second least among these approaches. 

\begin{figure}[htb]
\centering
\subfigure[]{
\begin{minipage}[t]{0.5\linewidth}
\centering
\includegraphics[width=4.0cm]{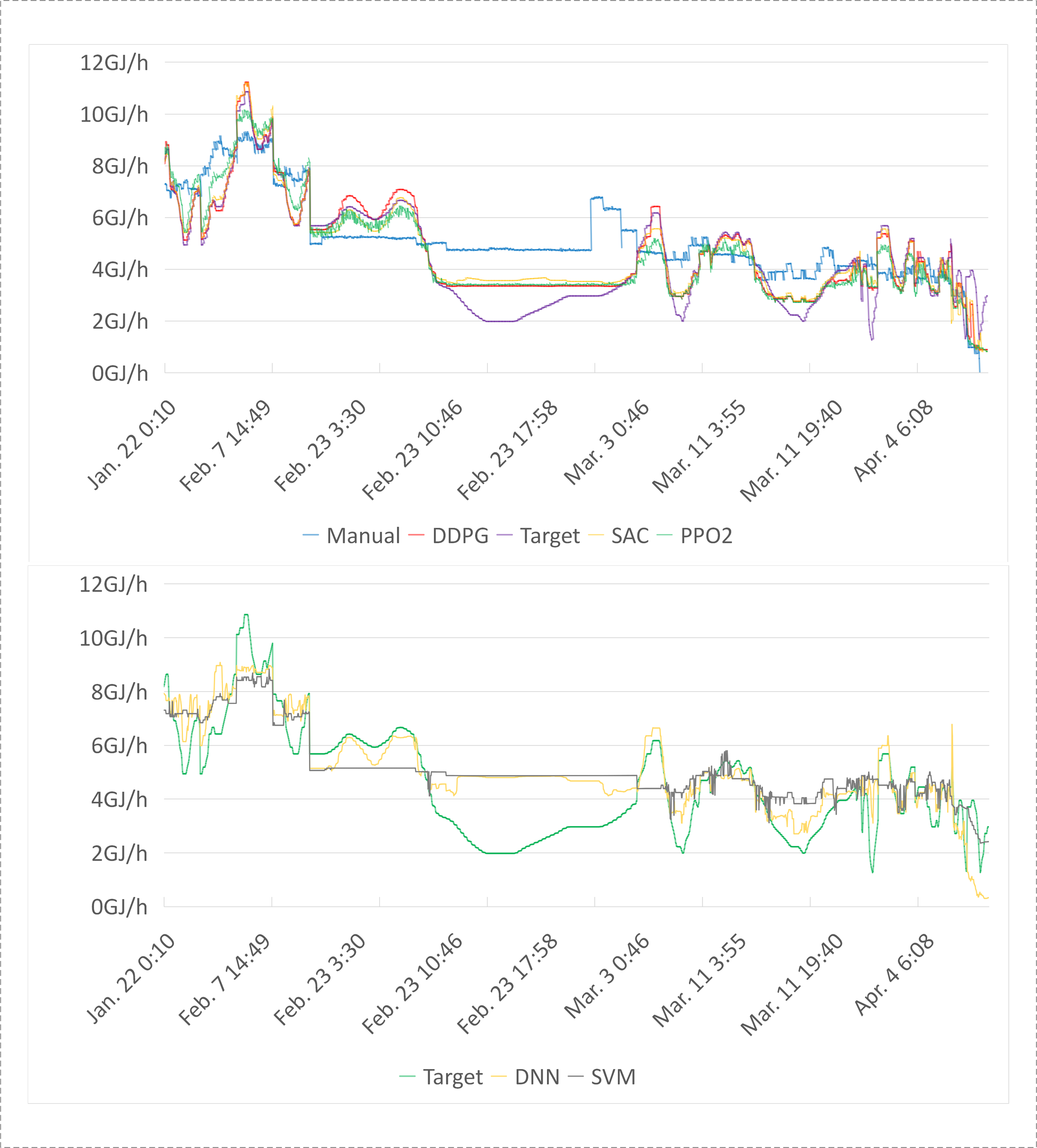}
\end{minipage}%
}%
\subfigure[]{
\begin{minipage}[t]{0.5\linewidth}
\centering
\includegraphics[width=4.0cm]{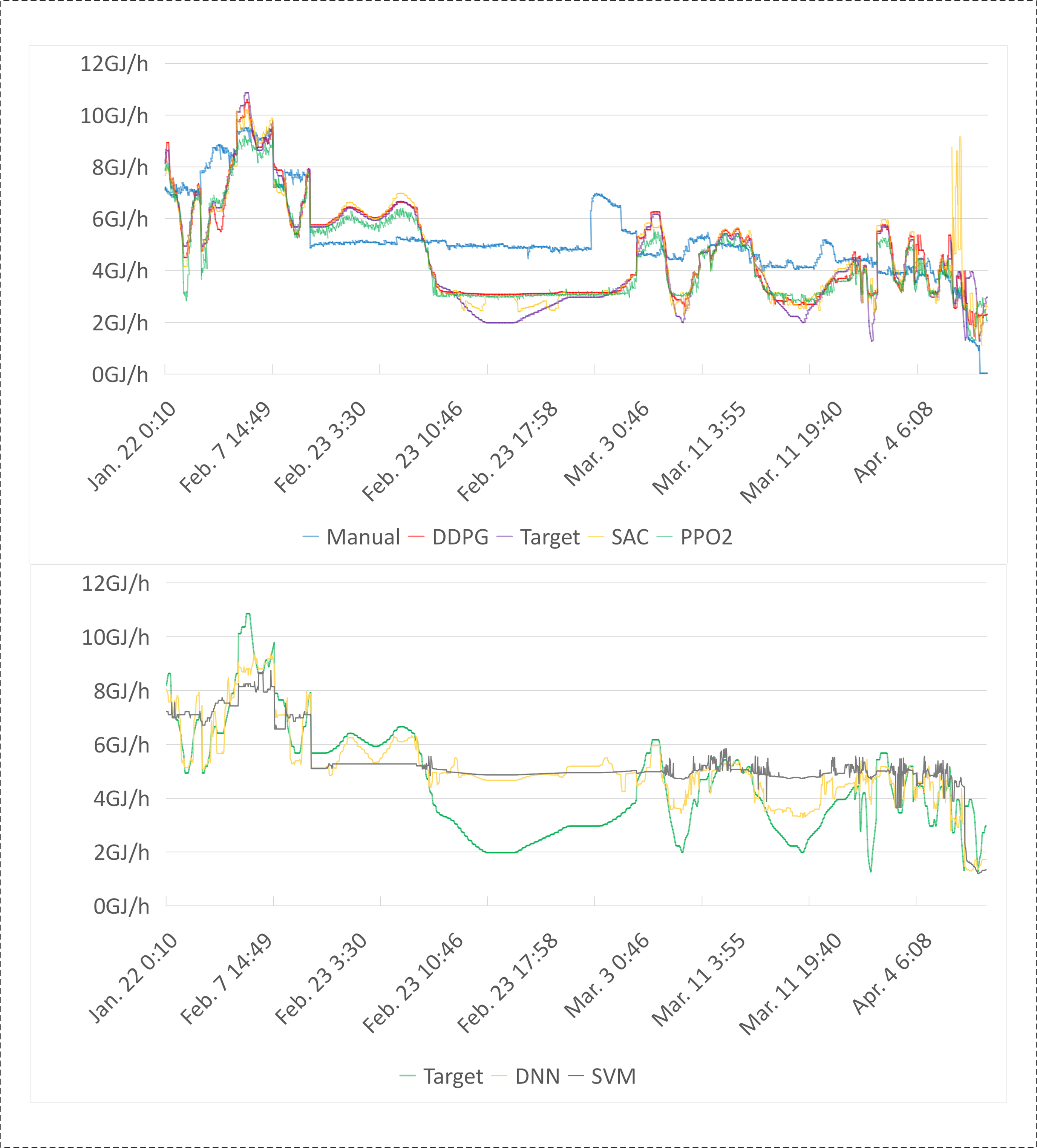}
\end{minipage}%
}%
\caption{ Daily Heat Quantity: (a) Primary Side; (b) Secondary Side.}
\label{meirire}
\end{figure}

For a more comprehensive comparison, the daily heat quantity is shown in Figure~\ref{meirire}. The curves closer to the target curve indicated a better performance. The DRL methods, particularly DDPG and SAC agents, stayed closer to the target curve, while the SL methods could not generate the appropriate heat quantity. Therefore, DRL was more suitable for DHS control. Moreover, we can also calculate the CE for each control method from Figure~\ref{meirire}, and the CE value of DDPG (1307.07 GJ/h) was the least. 

\begin{figure}[htb]
\centering
\subfigure[]{
\begin{minipage}[t]{0.5\linewidth}
\centering
\includegraphics[width=4.0cm]{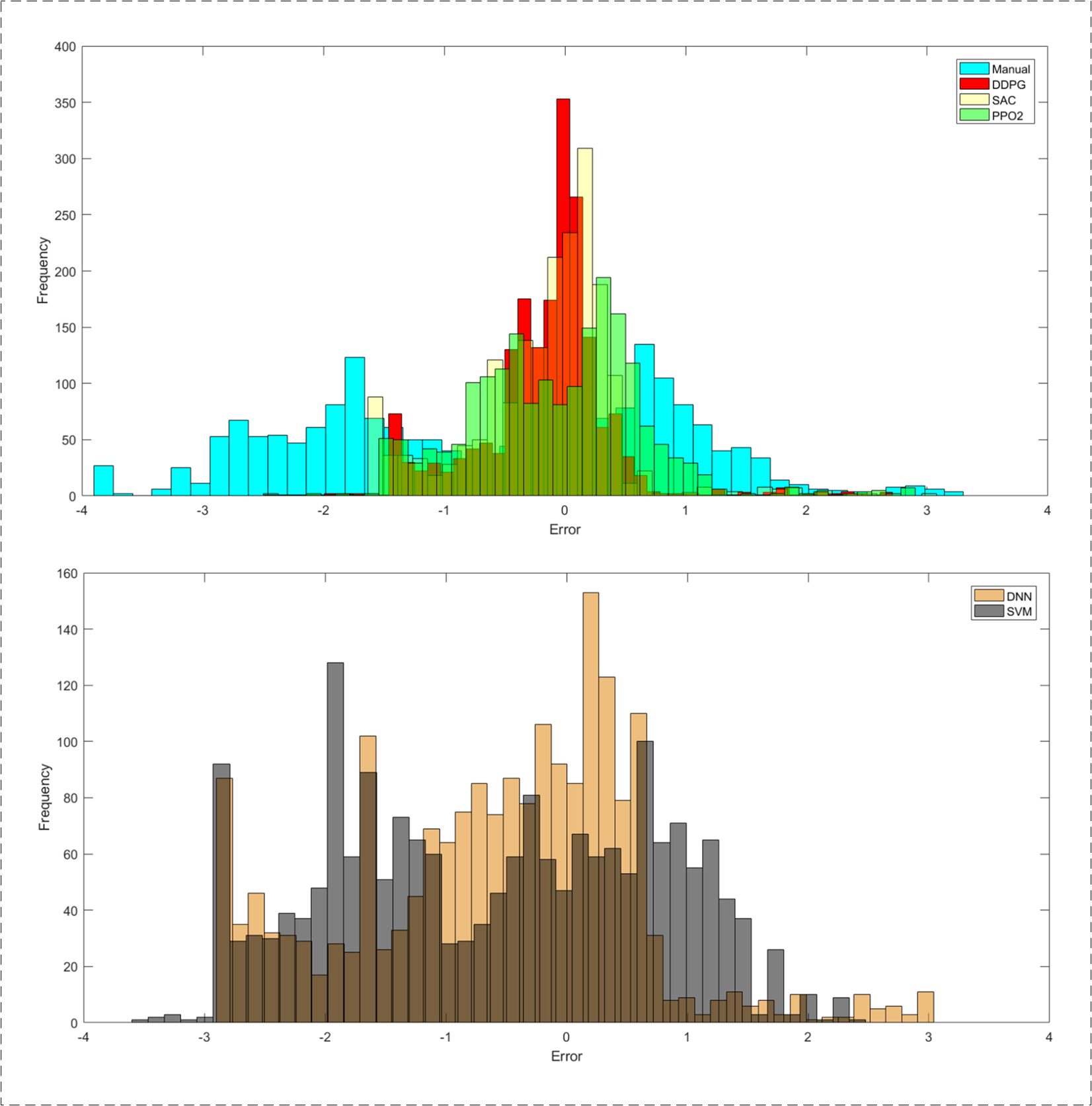}
\end{minipage}%
}%
\subfigure[]{
\begin{minipage}[t]{0.5\linewidth}
\centering
\includegraphics[width=4.0cm]{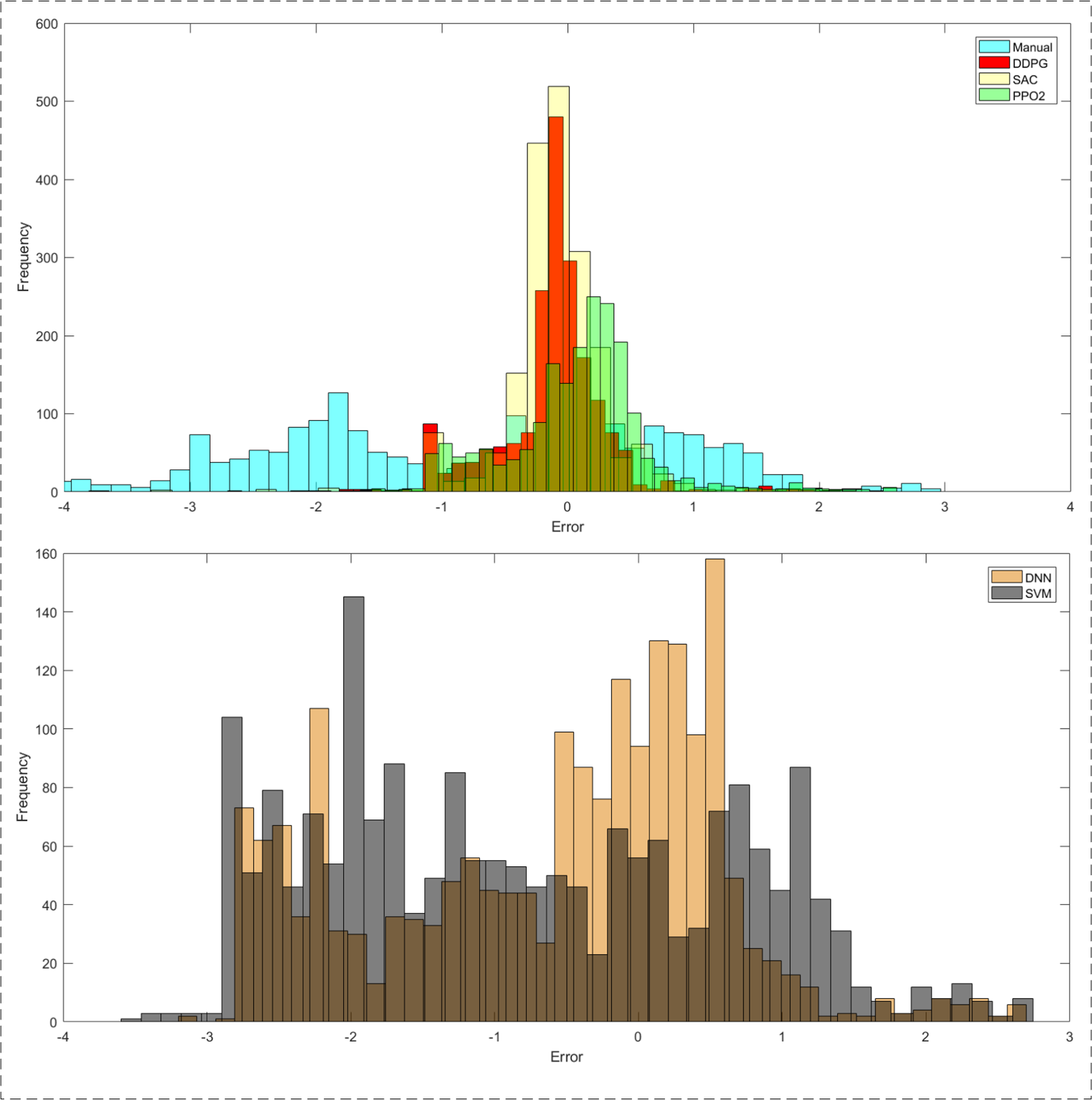}
\end{minipage}%
}%
\caption{Frequency Distribution Histogram of Heat Quantity Errors: (a) Primary Side; (b) Secondary Side.}
\label{pinlv}
\end{figure}

In order to intuitively observe the performance gaps among these methods, we calculated the errors between the target and the real heat quantity for each sample in the testing set, and plotted a frequency distribution histogram, as shown in Figure~\ref{pinlv}. The result showed that the errors of DDPG and SAC control stayed considerably close to zero, hence indicating that these were the most precise control method. More specifically, DDPG control performed better than SAC control on the primary side but not as well as SAC control on the secondary side. Considering other factors such as the water and heat energy consumption, we will adopt DDPG for the next deployment. In addition, the histogram significantly proved that the SL method could not perform well for the flow rate control in DHS. 

\textbf{Real-World Implementation for DDPG.} If operators plan to control a DHS by DRL for the first time, the lack of relevant data is the main problem for implementation. Therefore, we propose a rolling training approach for a DRL application to real-world problems.  

\begin{figure}[htb]
\centering
\includegraphics[width=8cm]{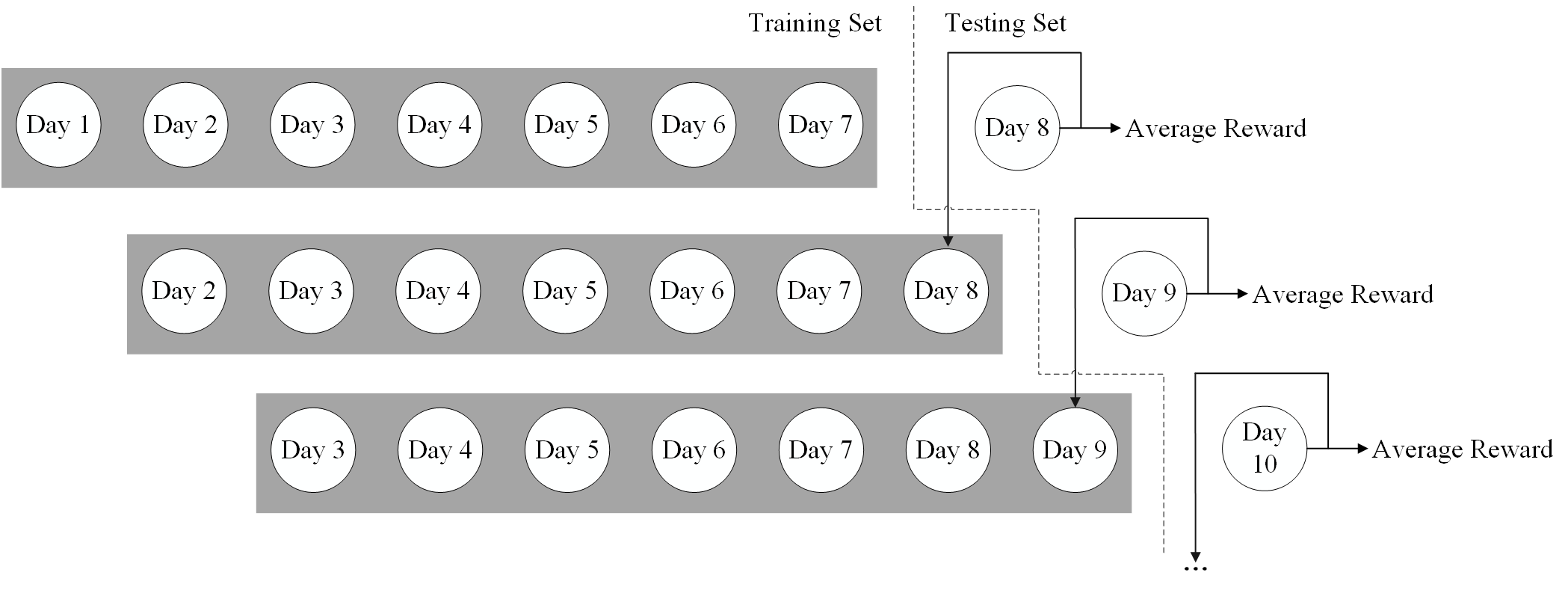}
\caption{Rolling Training for DDPG} 
\label{fig:roll}
\end{figure}

\begin{figure}[htb]
\centering
\includegraphics[width=8cm]{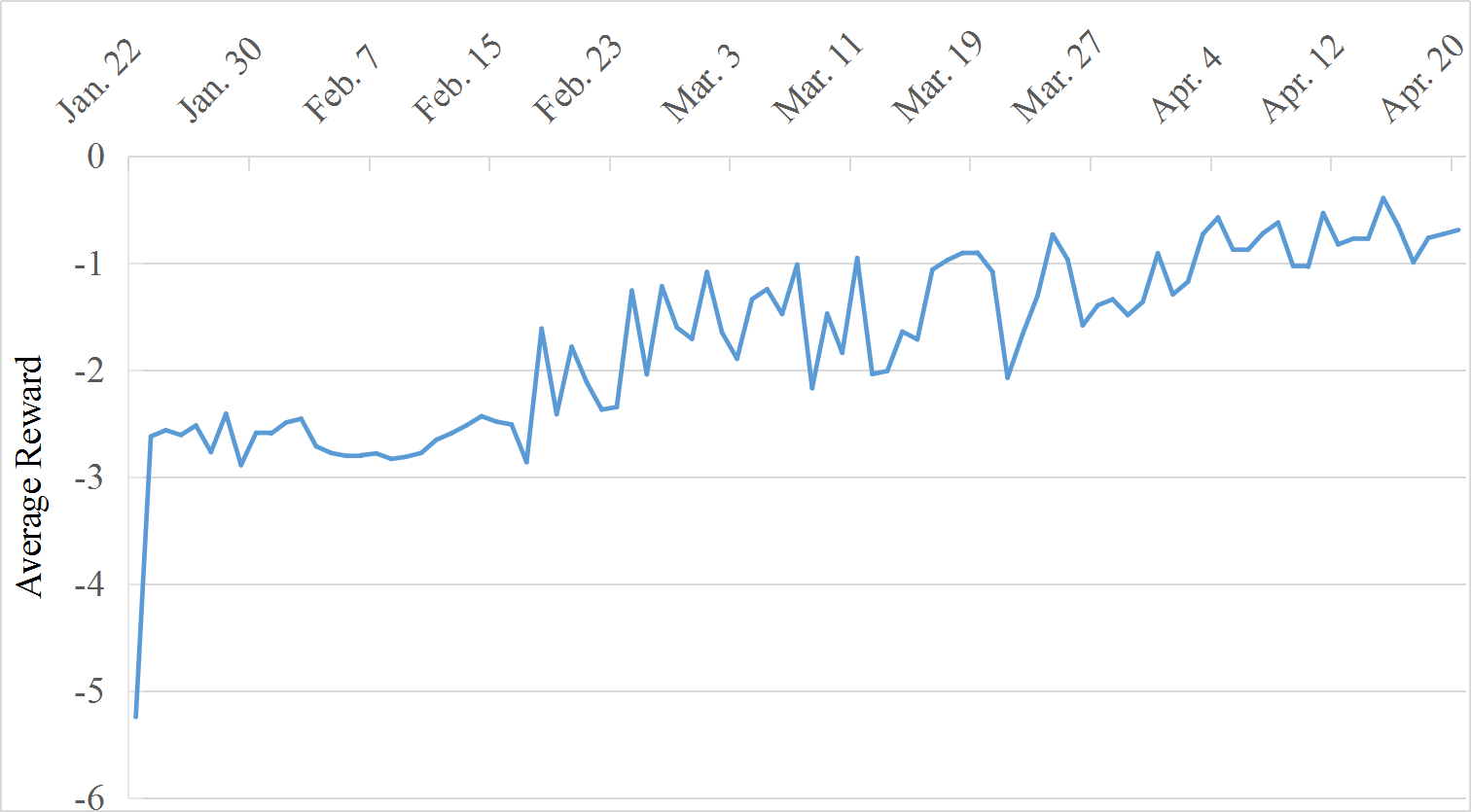}
\caption{Average Reward of Rolling Training} 
\label{fig:rollr}
\end{figure}

To begin with, the mechanism of rolling training is shown in Figure~\ref{fig:roll}. We first used the first seven-day data to train the agent, and the data of the eighth day were regarded as the testing set to evaluate the model. Then, the data from the second day to the eighth day were regarded as the training set, and those of the ninth day were the testing data, and so on. For saving time, each training process only had 20000 steps, but DDPG already provided good performance, as shown in Figure~\ref{fig:rollr}. As the number of samples for the daily data recording was different because of the unfixed sampling intervals, we calculated the average reward (AR) of the testing set as equation~\ref{kkkk} instead of the total rewards for the performance evaluation of DDPG,

\begin{equation}\label{kkkk}
        AR = \Sigma^M_{i}\frac{{\vert Q^i_1-Q^i_{target}\vert+\vert Q^i_2-Q^i_{target}\vert)}}{2M},
\end{equation}

where M denotes the number of samples in the testing set. On the first few days, DDPG was still learning and not good at the flow rate control. With the passage of time, it became more experienced and the AD value kept increasing, which meant that the agent was getting better.

\section{Conclusions}
In this study, we adopted deep reinforcement learning to deal with heating control problems. DDPG managed to regulate the flow rates of both sides to reach the target heat quantity calculated according to the national standard. Moreover, we developed a heating control system for real-world implementations. In particular, the heating parameters were collected and transferred through NB-IoT to Alibaba Cloud. Then, the balance controllers with the PID algorithm in the apartment and the remote centralized control for the flow rate worked together to regulate the entire heating system. A real-world case study was provided to show the system's practicability and the effectiveness of the PID controller; we also conducted a simulation experiment that proved that DDPG could exhibit better performance than manual control. 

However, there may be a time delay in using the DDPG control in real-world DHS control, as the heating process in a room is relatively slow. In the future, we can calculate the target heat quantity by using the outdoor temperature a few hours later, according to the weather forecast. Furthermore, there are still some other ways to apply DRL without using the target heat quantity. For example, researchers can install indoor temperature sensors for some individual users in an apartment. The reward function could be the average difference between the real room temperature (or the room temperature a few hours later) and the target temperature $T_0$.

\footnotesize
\bibliographystyle{acm}
\bibliography{sample-base}
\end{document}